\documentclass[%
 aip,
 amsmath,amssymb,
 reprint,%
]{revtex4-1}

\usepackage{graphicx}
\usepackage{dcolumn}
\usepackage{bm}

\usepackage[utf8]{inputenc} 
\usepackage[T1]{fontenc} 
\usepackage{mathptmx} 
\usepackage{etoolbox}
\usepackage{braket}    
\usepackage{nicefrac}
\usepackage{chemfig}

\usepackage{rotating}
\usepackage{makecell}
\usepackage{multirow}
\usepackage{array}
\usepackage{siunitx}
\usepackage[version=3]{mhchem} 
\usepackage{tikz}
\usepackage{tikz}
\tikzset{
  level/.style   = { ultra thick, black },
  connect/.style = { dashed, gray },
  notice/.style  = { draw, rectangle callout, callout relative pointer={#1} },
  label/.style   = { text width=2cm }
}

\newcommand{\lml}{ \ket{l,m_l}}

\newcommand{\JMJ}{ \ket{J,M_J}}
\newcommand{\LML}{ \ket{L,M_L}}

\newcommand{\LMLSMS}{ \ket{L,M_L,S,M_S}}

\newcommand{\tens}[1]{%
  \mathbin{\mathop{\otimes}\limits^{#1}}%
}

\makeatletter
\def\@email#1#2{%
 \endgroup
 \patchcmd{\titleblock@produce}
  {\frontmatter@RRAPformat}
  {\frontmatter@RRAPformat{\produce@RRAP{*#1\href{mailto:#2}{#2}}}\frontmatter@RRAPformat}
  {}{}
}%
\makeatother
\begin{document}

\preprint{AIP/123-QED}

\title{The NewMag crystal-field code for f-element systems: Implementation for extended active spaces, second-order correlated energies, and generalisation to $\bm{f^n}$ configurations }

\author{Gwenhaël Duplaix-Rata}
\affiliation{Univ Rennes, CNRS, ISCR (Institut des Sciences Chimiques de Rennes) – UMR
6226, F-35000 Rennes, France}

\author{Dumitru-Claudiu Sergentu}
\affiliation{Faculty of Chemistry, ''Alexandru Ioan Cuza'' University of Ia\textcommabelow{s}i, Ia\textcommabelow{s}i, Romania} 
\affiliation{ICI-RECENT AIR (RA03), ''Alexandru Ioan Cuza'' University of Ia\textcommabelow{s}i, Ia\textcommabelow{s}i, Romania}

\author{Boris Le Guennic}
\affiliation{Univ Rennes, CNRS, ISCR (Institut des Sciences Chimiques de Rennes) – UMR
6226, F-35000 Rennes, France}

\author{R\'emi Maurice*}
\affiliation{Univ Rennes, CNRS, ISCR (Institut des Sciences Chimiques de Rennes) – UMR
6226, F-35000 Rennes, France}

\email{remi.maurice@univ-rennes.fr}

\date{\today}


\begin{abstract}
We report a massive update of the \texttt{NewMag} program, which enables to extract Stevens crystal-field parameters (CFPs) after relativistic and multiconfigurational calculations are performed. The code can now post-treat ORCA and OpenMolcas outputs that contain state-average complete active space self-consistend field (SA-CASSCF) calculations with minimal or extended active spaces, second-order NEVPT2 or CASPT2 calculations, and spin-orbit configuration interaction  (SOCI) calculations. Stevens parameters are extracted following the original Stevens convention with the so-called ``extended'' parameters. Rotation invariant indicators are also computed, after applying a correction of the parameter values to ensure normalization, following Rudowicz. At the SOCI level, the spin-orbit coupling (SOC) constant is also extracted within a spherical approximation of the SOC operator. For all the reference calculation levels, the model spectrum is reconstructed, allowing a direct assessment of its quality. The reported implementation is successfully applicable to $f$-element systems with a non-void, non-half-filled/empty or non-full $f$ shell, that is for $f^n$ configurations with $n\neq$ 0, $\neq$ 7 and $\neq$ 14. Similarities and differences with the \texttt{SINGLE\_ANISO} and \texttt{AILFT} codes are discussed. With selected examples, we  showcase the interest of determining CFPs in $f$-element systems to understand their magnetic and optical properties in general, and more specifically the added value of \texttt{NewMag}. Finally, application of this approach to $d$-element systems is discussed, to reveal when it readily and successfully applies and when it may fail in reproducing satisfactorily the \textit{ab initio} energies. 
\end{abstract}

\maketitle

\section{\label{sec:introduction}Introduction\protect\\ }
The crystal field (CF) and ligand field (LF) models are key to understand various properties of $f$-element complexes, in particular concering the lanthanides (4$f$). The distinction between the CF and LF models is not always clearly articulated in the literature. In principle, however, the ``crystal field'' describes a metal ion surrounded by point charges, whereas the ``ligand field'' calls explicitly for actual ligands \cite{Chilton:2022} and modeling of their own electron-orbitals. In any case, these models deal with the valence atomic orbitals of the metal, such as the 4$f$ orbital shell of a given lanthanide. The aim is to describe the actual or effective orbital splitting and mixing induced by the presence of a set of point charges (CF models) or ligands (LF models), either at the one-electron orbital level (monoelectronic picture) or at the many-electron orbital level (polyelectronic picture). Unless specified otherwise, note that we will refer to the CF model thoughout this article without properly distinguishing the CF and LF models, as done by others \cite{Chilton:2022}. 

Actually, optical and magnetic properties of $f$-element systems can successfully be understood within the framework of CF theory \cite{Chilton:2022,Chilton:2025}. For instance, a ``good'' single ion magnet (SIM) may be obtained by following specific design rules that are directly related to the CF model \cite{Rinehart}. Regarding optical properties, the CF theory may allow to understand why formally-forbidden electronic transitions \cite{SlectionRules_Martin2006} become allowed, often quite intense, and may be observed in the laboratory. In both cases, one needs to introduce at least one orbital or one orbital+spin basis, and thus, we must introduce at this stage more specifically our CF framework. Note that various frameworks coexist in the literature and that we are working within the framework of Stevens equivalent operators \cite{Stevens:1952a}. The general expression of the CF Hamiltonian, $\hat{H}_{CF}$, is: 
\begin{equation}
\hat{H}_{CF}= \sum_{k}\alpha_{k}(J,L,l) \sum_{q=-k}^k B_{k}^{q} O_{k}^{q}(J,L,l)
\end{equation}
\noindent where the $k$'s and $q$'s are the operator ranks and orders, \mbox{respectively}, the $O_{k}^{q}$'s are the ``extended'' Stevens operators, the $B_{k}^{q}$'s are the CF parameters, and the $\alpha_{k}$'s are the appropriate reduced-matrix elements (these are also called $\alpha$, $\beta$ and $\gamma$ for rank 2, 4 and 6). $J$, $L$, $l$ are the possible bases: $J$ is the polyelectronic orbital+spin basis ($J$=$L$+$S$), $L$ is the polyelectronic orbital basis and $l$ is the monoelectronic orbital basis. For describing the magnetic properties of lanthanide SIMs, it is common practice to work with the ground-state $|J,M_J\rangle$ basis. A potentially ``good'' SIM may be expected if the $|J,\pm M_J^\text{max}\rangle$ energy levels are the lowest in energy, which translates into the oblate \textit{vs.} prolate rule of Rinehart and Long \cite{Rinehart}. 

The CF theory has undergone various conceptual developments over the past century and continues to evolve today, particularly within the computational chemistry community. Diverse CF programs have been developed, such as SIMPRE \cite{SIMPRE_Baldovi:2013a}, PyCrystalField \cite{PyCrystalField_Scheie:in5044}, NJA-CFS \cite{NJACFS_Fiorucci:2025}, \texttt{SINGLE\_ANISO} \cite{SingleAniso_Chibotaru:2012a}, and \texttt{AILFT} \cite{ailft_Atanasov:2015a, esAILFT}. The latter two are clearly the closest ones to our own work with \texttt{NewMag} \cite{NewMag_Sergentu:2026}, since these pioneered the extraction of CF parameters from relativistic and multiconfigurational \emph{ab initio} calculations, and have, of course, been a great source of inspiration. Note that the other mentioned programs typically execute an empirical extraction using a point-charge model to represent the ligands. 

The \texttt{SINGLE\_ANISO} program, developed by Ungur and Chibotaru \cite{SingleAniso_Chibotaru:2012a,SingleAniso_Ungur:2017a}, was the first available for extracting CF parameters from relativistic and multiconfigurational \emph{ab initio} wave functions, and played a key role in the development of the field of computational magnetism with transition-metal and lanthanide complexes; the code allowed many researchers to rationalize magnetic properties across a plethora of SIMs, and contributed decisively to the design and characterization of current state-of-the-art lanthanide-based SIMs. This program works with the $L$ and $J$ bases. With the $L$ basis, all components of the ground spin-orbit-free terms of the reference free ion are retained (2$L$+1 components). With the $J$ basis, only the $M_J$ components of the ground $^{2S+1}L_J$ term are included (2$J$+1 components). 

The \texttt{AILFT} (for \emph{Ab Initio} Ligand Field Theory) program, developed by Atanasov \emph{et al.} \cite{ailft_Atanasov:2015a}, is more recent. It considers the $l$ basis to extract information regarding the CF, and computes the spin-orbit coupling (SOC) constant, $\zeta_\text{SOC}$, through a fitting of the corresponding ligand-field SOC Hamiltonian to the \emph{ab initio} SOC matrix. Interestingly, this program can also compute Slater-Condon parameters which aim at accounting for the electron-electron term of the atomic Hamiltonian. Naturally, both these programs have their own advantages and disadvantages. 


The ``pseudo-spin'' approximation is adopted by the \texttt{SINGLE\_ANISO} code \cite{SingleAniso_Chibotaru:2012a}, in fact the pseudo-$L$ or pseudo-$J$ approximation for the sake of the present article. The first 2$L$+1 roots at the scalar-relativistic (SR) level are supposed to correlate with the 2$L$+1 components of the ground $^{2S+1}L$ term of the reference free ion, and the first 2$J$+1 roots of the spin-orbit configuration interaction (SOCI) calculation are supposed to correlate with the 2$J$+1 components of the ground $^{2S+1}L_J$ term. If that is not the case, as it is often encountered with actinides in particular \cite{Sergentu:2018}, the data generated by the code becomes meaningless. Moreover, the $f^6$ configuration remains unexplored at the SOCI level, within a pseudo-$J$ formalism, simply because the ground manifold of such a free ion only consists of one orbital+spin configuration ($J=0$), hence there is no CF splitting within this space and thus no CF parameters (CFPs) can be extracted from it. Of course, the CFPs are accessible at the SR level, within the pseudo-$L$ approximation, and this is not new, it was already possible with \texttt{SINGLE\_ANISO}.

The \texttt{AILFT} code is somehow more general, since it also computes Slater-Condon parameters and the SOC constant (within the spherical approximation). Furthermore, it treats the $f^6$ configuration on the same footing as the other $f^n$ cases, apart from the $f^0$ and f$^{14}$ configurations, which are irrelevant for the CF, and the $f^7$ configuration, which is a special case in its own right, since there is no first-order SOC contribution within the ground $^{2S+1}L_J$ term of the reference free ion when $L=0$. Extraction of the CF at the monoelectronic orbital level ($l$ basis) triggers both an advantage and a disadvantage: on the one hand, it is more general than \texttt{SINGLE\_ANISO} in terms of the $f^n$ configurations that can be tackled; on the other hand, apart from specific configurations (namely the $f^1$, $f^6$, $f^8$ and $f^{13}$ configurations), it ultimately leads to different CFP values as the ones that may be obtained based on polyelectronic bases, be them related to $L$ (SR level) or $J$ (SOCI level).
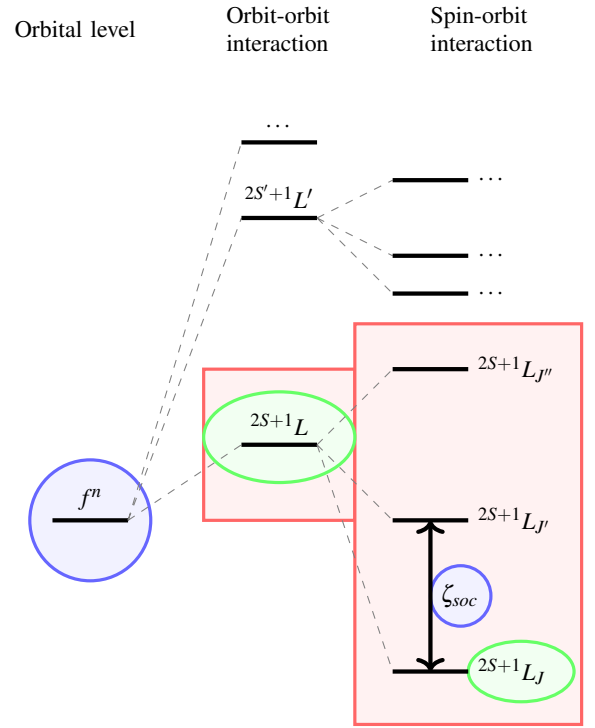
\begin{figure}
\begin{tikzpicture}
  \filldraw[color=red!60, fill=red!5, very thick](2,0) rectangle ++(2, 2);
  \filldraw[color=red!60, fill=red!5, very thick](4,-2.7) rectangle ++(3, 5.3);

  \filldraw[color=blue!60, fill=blue!5, very thick](0.5,0) ellipse (0.8 and 0.8);
  \filldraw[color=blue!60, fill=blue!5, very thick](5.4,-1) ellipse (0.4 and 0.4);

  \filldraw[color=green!60, fill=green!5, very thick](3,1.1) ellipse (1 and 0.6);
  \filldraw[color=green!60, fill=green!5, very thick](6.2,-2) ellipse (0.7 and 0.4);

  \draw[level] (0,0) -- node[above] {$f^n$} (1,0);

  \draw[connect] (1,0)  -- (2.5,1) (1,0)  -- (2.5,4) (1,0)  -- (2.5,5) ;
  \draw[level]   (2.5,1)  -- node[above] {$^{2S+1}L$} (3.5,1);
  \draw[level]   (2.5,4)  -- node[above] {$^{2S'+1}L'$} (3.5,4);
  \draw[level]   (2.5,5)  -- node[above] {$\cdots$} (3.5,5);
  
  \draw[connect] (3.5,1)  -- (4.5,-2) (3.5,1)  -- (4.5,0)  (3.5,1) -- (4.5,2);
  \draw[level]   (4.5,-2) -- (5.5,-2) node[right] {$^{2S+1}L_{J}$};
  \draw[level]   (4.5,0)  -- (5.5,0) node[right] {$^{2S+1}L_{J'}$};
  \draw[level]   (4.5,2)  -- (5.5,2) node[right] {$^{2S+1}L_{J''}$};

  \draw[level, <->]   (5,-2) -- (5,0) node[midway,right] {$\zeta_{soc}$};

  \draw[connect] (3.5,4)  -- (4.5,3) (3.5,4) -- (4.5,3.5) (3.5,4) -- (4.5,4.5);
  \draw[level]   (4.5,3) -- (5.5,3) node[right] {$\cdots$};
  \draw[level]   (4.5,3.5) -- (5.5,3.5) node[right] {$\cdots$};
  \draw[level]   (4.5,4.5) -- (5.5,4.5) node[right] {$\cdots$};

  \node[label] at (0.5,6.5)  {Orbital level};
  \node[label] at (3.3,6.5)  {Orbit-orbit interaction};
  \node[label] at (6,6.5) {Spin-orbit interaction};
\end{tikzpicture}
\label{Energy_level_diagrams}
\caption{Generic diagram of energy levels for an $f^n$ configuration, illustrating the areas of application of \texttt{SINGLE\_ANISO} \cite{SingleAniso_Ungur:2017a} (green ellipsoids), \texttt{AILFT} \cite{ailft_Atanasov:2015a} (blue circles), and \texttt{NewMag} \cite{NewMag_Sergentu:2026} (red rectangles).}
\end{figure}

Alternatively, we have recently reported \texttt{NewMag}: a new code to compute parameters for understanding optical and magnetic properties in the context of \emph{ab initio} multiconfigurational calculations \cite{NewMag_Sergentu:2026, NewMag:version1}. Figure \ref{Energy_level_diagrams} illustrates common points and differences between \texttt{NewMag} and other codes. At the SR level, \texttt{NewMag} is very close to \texttt{SINGLE\_ANISO} (same basis, same target), and we have developed a variant that combines features of both \texttt{SINGLE\_ANISO} and  \texttt{AILFT} at the SOCI level: the SOC constant is extracted under the spherical approximation, as in \texttt{AILFT}, whereas the CFPs are extracted in a manner similar to that used in \texttt{SINGLE\_ANISO}, but using a different basis: the 2$J$+1 components of all the $J$ terms correlating with the ground $^{2S+1}L$ term of the reference free ion, instead of only the 2$J$+1 components of only one $^{2S+1}L_J$ term, as in \texttt{SINGLE\_ANISO}. 

The first \texttt{NewMag} implementation treated the $f^1$ configuration only, based on minimal active space calculations \cite{NewMag_Sergentu:2026}. In this work, the code is extended to allow for the use of larger active spaces, root selection beyond the pseudo-$L$ or pseudo-$J$ approximation, diagonally dressed SOCI calculations (with ``correlated'' energies), and is generalized to modeling all the $f^n$ configurations of interest (\textit{i.e.} $f^2$--$f^6$,$f^8$--$f^{13}$). \texttt{NewMag} can now process both ORCA \cite{Orca_Neese:2025a} and OpenMolcas \cite{OpenMolcas_Manni:2023a} outputs, and extracts CF Hamiltonians at the state-averaged complete active space self consistent-field (SA-CASSCF) level \cite{CASSCF_Roos:1980a,CASSCF_Roos:1980b}, at the correlated CASPT2 \cite{ss_CASPT2_Andersson:1990a,ss_CASPT2_Andersson:1992a} and NEVPT2 \cite{NEVPT2_Angeli:2001a} levels, and the CF + SOC Hamiltonian at the SOCI level \cite{RASSI_Malmqvist:2002a,Roos:2004}. Extending CF extractions beyond the minimal valence shell is also a current goal of the \texttt{AILFT} family: Rao \textit{et al.}\ recently introduced an extended-active-space \texttt{AILFT} (esAILFT) in which the configurations outside the minimal $d$ (or $f$) space are down-folded into the minimal-space CI matrix by L\"owdin partitioning\cite{esAILFT}. This preserves the parametrization and has so far been benchmarked on 3$d$--5$d$ free ions. In contrast, \texttt{NewMag} works in the many-electron $L$/$J$ multiplet basis and projects the selected \textit{ab initio} states onto the model space, to extract Stevens CFPs. 

\texttt{NewMag} is intended as an alternative to \texttt{SINGLE\_ANISO} and  \texttt{AILFT}, combining their main advantages and hopefully limiting extra side effects. By selected examples, the manuscript demonstrates that \texttt{NewMag} is always successful with lanthanide complexes, and further discusses its applicability to transition metal and actinide complexes. 

\section{\label{sec:method}Method and implementation\protect\\ }

\subsection{\label{subsec:spaces}On model spaces \protect\\ }

Prior to presenting the core methodology, we first discuss the model spaces considered by \texttt{SINGLE\_ANISO}, \texttt{AILFT}, and \texttt{NewMag}. For a high-spin $f^n$ configuration (Hund's rule), the ground state of the free ion is $^{2S+1}L$, 2$S$+1 being the spin multiplicity and $L$ denoting the (total) orbital momentum, in other word the orbital degeneracy of this ground state, S for 1, P for 3, D for 5, \textit{etc.} At the SR level, both \texttt{SINGLE\_ANISO} and  \texttt{NewMag} work in the basis of $L$ (2$L$+1 roots). Therefore, the number of roots depends on the configuration, with a symmetry between the $f^1$--$f^6$ and $f^8$--$f^{13}$ series (see Table \ref{roots}). Inside these two series, additional symmetry is observed (between the $f^1$--$f^3$ and $f^4$--$f^6$ subseries, as well as between the $f^8$--$f^{10}$ and $f^{11}$--$f^{13}$ ones, respectively). Note that since \texttt{AILFT} works on the basis of $l$, the model space size remains constant (2$l$+1=7 for an $f$ shell).

\begin{table}[h!]
\caption{Number of spin-orbit free states within the pseudo-$L$ approximation, 2$L$+1, of energy levels within the pseudo$-J$ approximation, 2$J$+1, and of energy levels used in this work, $\sum\limits_J$2$J$+1. \\}
\label{roots}
\centering
\begin{tabular}{|c|cccccc|cccccc|}
\hline
$f^n$ & $f^1$  & $f^2$  & $f^3$  & $f^4$  & $f^5$  & $f^6$  & $f^8$  & $f^9$  & $f^{10}$  & $f^{11}$  & $f^{12}$  & $f^{13}$  \\
\hline
2$L$+1 & 7 & 11 & 13 & 13 & 11 & 7 & 7 & 11 & 13 & 13 & 11 & 7 \\
\hline
2$J$+1 & 6 & 9 & 10 & 9 & 6 & 1 & 13 & 16 & 17 & 16 & 13 & 8 \\
$\sum\limits_J$2$J$+1 & 14 & 33 & 52 & 65 & 66 & 49 & 49 & 66 & 65 & 52 & 33 & 14  \\
\hline
\end{tabular}
\end{table}

Once the SOC is considered, the model space size naturally increases. With \texttt{NewMag}, since it works on the basis of all the possible $J$'s that can be formed from the ground $^{2S+1}L$ term of the reference free ion, the model space includes all the spin components of the 2$L$+1 orbital roots, that is $\sum\limits_J$2$J$+1 = (2$L$+1)(2$S$+1) = (2$L$+1)($\tilde n$+1) where $\tilde n$ is the (maximum) number of unpaired electrons ($\tilde n$=$n$ if $n$<7 or $\tilde n$=14$-n$ if $n$>7). The spin thus breaks the symmetry between the $f^1$--$f^3$ and $f^4$--$f^6$ subseries, as well as between the $f^8$--$f^{10}$ and $f^{11}$--$f^{13}$ ones, while maintaining the symmetry between the $f^1$--$f^6$ and $f^8$--$f^{13}$ series (see Table \ref{roots}). With \texttt{SINGLE\_ANISO}, the pseudo-$J$ approximation implies that the ground $^{2S+1}L_J$ term of the reference free ion is used to define the model space. For a less than half-filled shell ($f^1$--$f^6$), $J^\text{min}$ is the ground state, while for a more than half-filled shell, $J^\text{max}$ is the ground state. This explains why the symmetry between the $f^1$--$f^6$ and $f^8$--$f^{13}$ series is broken in this case. Moreover, as already mentioned, the pseudo-$J$ approximation for the $f^6$ configuration leads to only one energy level in the model space, and thus to the practical impossibility of extracting CFPs. 

To conclude, prior to any extraction of the CFPs with \texttt{NewMag}, the previous SR \textit{ab initio} calculations must deal with at least 2$L$+1 roots while the SOCI ones must deal with at least (2$L$+1)($\tilde n$+1) roots, since by definition the \textit{ab initio} space cannot be smaller than the model one. We recall here that \texttt{NewMag} is a post-treatment code, requiring the use of a reference \textit{ab initio} code to perform the actual relativistic and multiconfigurational calculations, as currently ORCA or OpenMolcas.

\subsection{\label{subsec:caract_wfn}On relativistic and correlated wave functions \protect\\ }

In principle, one must feed \texttt{NewMag} with a set of at least 2$L$+1 SR roots and at least (2$L$+1)($\tilde n$+1) SOCI roots. In both cases, the wave functions are  multiconfigurational. Assuming SA-CASSCF calculations, the SR wave functions are typically expressed in the following way:

\begin{equation}
\psi^{\text {SR}}_{i} = \sum\limits_{l} a_{l}\left|\phi_{l}\right>
\label{eq:CAS}
\end{equation}

\noindent where each root $\psi^{\text {SR}}_{i}$ is expressed as a linear combination of Slater determinants $\phi_{l}$ with real coefficients $a_{l}$. Within each Slater determinant, the molecular (or atomic, for free atoms or ions) orbitals may have occupations of 2, 1 or 0, with the total number of electrons in a given Slater determinant equal to $n$. In a minimal active-space calculation for an $f^n$ configuration, this corresponds to $n$ electrons distributed among the seven $f$ orbitals.

The SOCI calculations, performed in a second step, generate roots that are expressed in terms of $M_S$ components of the previous SR states:

\begin{equation}
\psi^{\text {SOCI}}_{j} = \sum\limits_{i,M_S} b_{(i,M_S)}\left|\psi^{\text {SR}}_{i},M_S\right>
\label{eq:SOCI}
\end{equation}

\noindent where each root $\psi^{\text {SOCI}}_{j}$ is expressed as a linear combination of the $M_S$ components of the previous Slater determinants ($\psi^{\text {SR}}_{i}$), this time with complex coefficients $b_{(i,M_S)}$. 

In principle, knowledge of the molecular (or atomic) orbitals and of the appropriate configuration interaction (CI) coefficients---$a_{l}$ at the SR level and both $a_{l}$ and $b_{(i,M_S)}$ coefficients at the SOCI level---fixes all the properties of the wave functions of interest. It is recalled here for the sake of pedagogy that orbital rotations within the active space modify accordingly the $a_{l}$ CI coefficients, if the CI is performed in the rotated orbital basis, and that the $b_{(i,M_S)}$ coefficients remain invariant under such rotations. Such operations should, in no way, affect the intrinsic properties of the resulting wave functions. They merely provide the freedom to rotate the orbitals into a basis that facilitates the analysis and interpretation of the wave functions, which is the approach adopted here.

If minimal-active-space SA-CASSCF calculations are performed, and if the CASCI space for the high-spin states spans only the orbital configurations that correlate with the $^{2S+1}L$ term of the reference free ion, as is the case for the $f^1$ configuration, the model space and the CASCI space have the same size. Under these conditions, the assignment of a given SR state is relatively straightforward. However, when extended-active-space calculations are performed, the model space is necessarily smaller than the CASCI space, and there is a greater possibility of assigning SR states to the model space that do not actually belong to it. In order to develop a general code capable of handling even such tricky situations, we chose to work with ``localized'' and ``purified'' active orbitals, which allow us to identify an interpret easily the SR wave functions. Because the SOCI wave functions are constructed directly from the SR ones, by simply introducing the corresponding spin configurations, these wave functions retain the same degree of clarity. This situation differs from esAILFT\cite{esAILFT}, where the model space is fixed by the number of electrons in the $d$ ($f$) shells and the outer-space configurations are folded into it by partitioning theory, with a bias correction and with root filtering applied during the CASSCF step.

As a post-treatment code, \texttt{NewMag} builds upon the capabilities of previously used quantum-chemistry packages, ORCA or OpenMolcas. These are first used to perform SA-CASSCF calculations and localize the active space orbitals, which are fed to subsequent CASCI calculations. With ORCA, localized orbitals can be generated with the ``actorbs'' keyword, which in fact generates the orbital set that is also used by \texttt{AILFT}, or manually, using the ``rotate'' directive in the \texttt{scf} block. With OpenMolcas, localized orbitals can be generated either using classical localization schemes, or manually using external codes such as \texttt{morot} \cite{Sergentu:morot}. Output examples are provided in the \texttt{NewMag} GitHub repository, with inputs being displayed at the top, and the interested reader may follow the same output preparation workflow if wished. Going forward, at this stage, we have well-defined sets of $\psi^{\text {SR}}_{i}$ and $\psi^{\text {SOCI}}_{j}$ multiconfigurational wave functions. Also, note that both ORCA and OpenMolcas express the Slater determinants in terms of orbitals represented by real spherical harmonics (RSHs). A similar orbital-preparation strategy (localization, followed by diagonalization of $\hat{L}_z$ on the metal-based orbitals) is also used in esAILFT\cite{esAILFT}.

\subsection{\label{subsec:SR_H}Building the effective Hamiltonian at the SR level\protect\\ }

This section presents the derivation of the effective Hamiltonian at the SR level, in the basis of the $M_L$ components of the ground $^{2S+1}L$ term of the reference free ion. This (2$L$+1)$\times$(2$L$+1) effective Hamiltonian must, by construction, reproduce the SR \textit{ab initio} energies, while its eigenfunctions must correspond to the \textit{ab initio} wave functions projected onto the model space \cite{Bloch:1958}. Apart from the \textit{ab initio} energies, which are obviously printed directly by ORCA and OpenMolcas in the outputs, the \textit{ab initio} wave functions, which are expressed in the basis of Slater determinants represented by RSHs, are processed further by \texttt{NewMag} in five steps:

\begin{itemize}
\item[1.] Selection of the roots that essentially develop on the model space (from the $i$ roots of the CASCI space to the 2$L$+1 roots of the model space). 
\item[2.] Retain the components of the corresponding wave functions that are developed on the targeted $f$ orbitals (\textit{i.e.}, project the $\psi^{\text{SR}}$ wave functions to obtain the $\tilde{\psi}^{\text{SR}}$ wave functions).
\item[3.] Apply a first transformation to express the $\tilde \psi^{\text {SR}}$ wave functions in the basis of Slater determinants represented by complex spherical harmonics (CSHs).
\item[4.] Apply a second transformation to express the $\tilde \psi^{\text {SR}}$ wave functions in the $\LML$ many-electron orbital basis.
\item[5.] Build the des Cloizeaux effective Hamiltonian \cite{Cloizeaux:1960a} in the $\LML$ basis.
\end{itemize} 

Steps 2 and 5 are standard for extracting parameters within the effective Hamiltonian framework and will not be detailed here. It is recalled that the projected wave functions are Löwdin orthogonalized \cite{Lowdin:1950a} prior to building the effective Hamiltonian, ensuring its Hermiticity by construction. Additional details and references are readily available elsewhere \cite{Suaud:2026}, including in the original release paper of \texttt{NewMag} \cite{NewMag_Sergentu:2026}. Step 1, in contrast, is a key feature of the present \texttt{NewMag} development, as it allows treatments beyond the pseudo-$L$ approximation (which is not possible with \texttt{SINGLE\_ANISO}). \texttt{NewMag} allows the user to choose among the following options:

\begin{itemize}
\item[a.] Retain the pseudo-$L$ approximation, whereby the code directly retains the first 2$L$+1 roots from the output. In practice, the high-spin states must first be computed. Among these, the retained states are the first 2$L$+1 roots in ascending order of energy. With this option selected, \texttt{NewMag} behaves as \texttt{SINGLE\_ANISO} at the SR level. This is expected to be operative for most of if not all the lanthanide complexes, especially if minimal active space calculations are performed.
\item[b.] The user provides \texttt{NewMag} an output from an independent SA-CASSCF calculation of the reference free ion, where the first 2$L$+1 roots must belong to the model space. The code then analyze the \textit{ab initio} wave functions of the complex and retains the 2$L$+1 roots with the largest projections onto the 2$L$+1 roots of the reference free ion.
\item[c.] The user manually specifies the roots to be retained. This option is particularly useful when the appropriate roots can be readily identified by inspecting visually the wave functions in the output; the localization described above is crucial for this purpose.
\end{itemize}

While the transformations described in Steps 3 and 4 are trivial for the $f^1$ configuration \cite{NewMag_Sergentu:2026}, they are somewhat more complex for other $f^n$ configurations. Indeed, in the $f^1$ configuration, transformation from the RSHs to the CSHs (Step 3) is simply done by applying a $U$ matrix \cite{RSH_CSH:1997a} and since this is a one-electron case, transformation to the many-electron orbital basis $\LML$ (Step 4) is unnecessary. To generalize, let us denote by $U_1$ the 7$\times$7 matrix that transforms the one-electron representation from the RSHs to the CSHs. In $f^n$ cases, the transformation matrix $U$ is constructed as the tensor product of $U_1$ with itself, repeated $\tilde n-1$ times:

\begin{equation}
U = U_1 \tens{\tilde n-1} U_1 = \underbrace{U_1 \otimes ... \otimes U_1}_{\tilde n-1}
\label{eq:u_matrix}
\end{equation}

\noindent Thus, the $U$ matrix is 7$^{\tilde n}\times$7$^{\tilde n}$ dimensional, \textit{i.e.} 7$\times$7 matrix in the $f^1$ and f$^{13}$ cases ($\tilde n$=1), 49$\times$49 matrix in the  $f^2$ and f$^{12}$ cases ($\tilde n$=2), \textit{etc.} up to 117649$\times$117649 in the $f^6$ and f$^{8}$ cases ($\tilde n$=6). 

The $U$ matrix may contain unnecessary configurations, \textit{i.e.} those that are not high-spin or incompatible with the Pauli exclusion principle. For instance, in the $f^2$ case, only 42 configurations may lead to high-spin configurations (2 unpaired electrons), while 7 configurations can only lead to closed-shell configurations. In the $f^3$ case, 210 configurations may lead to high-spin configurations (3 unpaired electrons), 126 configurations may only lead to low-spin configurations (only 1 unpaired electron) and 7 configurations are not compatible with the Pauli exclusion principle (3 electrons occupying the same orbital). It is clear that Equation \eqref{eq:u_matrix} necessarily generates unnecessary configurations; however, their inclusion is required for the transformation to remain consistent. The $\tilde \psi^{\text {SR}}$ wave functions must then be expressed in the same basis and with the same ordering as $U$. This requires their expansion to be extended to include the unnecessary configurations, with the corresponding coefficients set to zero. This procedure is fully automated in \texttt{NewMag} and therefore requires no intervention from the user.

In a final step, a transformation matrix $C$ constructed with Clebsch-Gordan coefficients, must be applied to reach the $\LML$ basis. Assuming $\tilde n-1$ successive couplings, $C$ is not only based on products of Clebsch-Gordan coefficients, since there are various ways to generate a given $\LML$ function by coupling the $\lml$ functions. Instead, appropriate products of Clebsch-Gordan coefficients must be summed for all possible patterns leading to a given $\LML$ function. Note that the wave functions are expressed with complex coefficients after both Steps 3 and 4, the CI coefficients being initially real only in the basis of the RSHs.

All the appropriate transformation matrices have been implemented in \texttt{NewMag}, and the code can now safely handle all the $f^n$ configurations of interest ($n$ = 1--6 and $n$ = 8--13). Although this step-by-step procedure may seem tedious, involving the localization of the orbitals following an initial SA-CASSCF calculation, a subsequent CASCI calculation, and several successive basis transformations, it is necessary to track the character of the many-electron states throughout the entire procedure, without introducing any \textit{a priori} assumptions regarding their nature. By doing so, one can extract CF Hamiltonians fully \textit{ab initio}; the approach may allow one to go beyond the phenomenological approach if required, since the \textit{ab initio} calculations can then support the choice of the model Hamiltonian or help revising it if necessary \cite{Suaud:2026}. This approach also goes beyond the pseudo-$L$ approximation, thereby making it more generally applicable than \texttt{SINGLE\_ANISO}.

\subsection{\label{subsec:SOC_H} Building the effective Hamiltonian at the SOCI level\protect\\ }

The SOCI wave functions are expressed in terms of $M_S$ components of the SR states. It is now straightforward to express these in terms of $\LMLSMS$ basis states since the SR states were already prepared in terms of the $\LML$ functions (vide supra). In lanthanide complexes, it is common to work in the ``coupled'' $\JMJ$ basis, especially if only first-order SOC is considered, as it is the case with the current implementation of \texttt{NewMag}. This is due to the fact that in lanthanide complexes, the CF is a much weaker perturbation of the atomic picture than the SOC is, or, in other words, to the fact that the Russel-Saunders coupling dominates. The following procedure is adopted to derive the effective Hamiltonian in the coupled basis:

\begin{itemize}
\item[1.] Selection of the roots that essentially develop on the model space (from the $j$ roots of the SOCI space to the $\sum\limits_J$2$J$+1 roots of the model space). \texttt{NewMag} simply computes the sum of the overlaps between each SOCI state and all the SR states belonging to the model space at the SR level. The sum should approach 100\% if the SOCI state of interest belongs to the model space.
\item[2.] Retain the parts of the corresponding wave functions that are only developed on the model space (projection, from $\psi^{\text {SOCI}}$ to $\tilde \psi^{\text {SOCI}}$ wave functions).
\item[3.] Express the $\tilde \psi^{\text {SOCI}}$ wave functions in the $\LMLSMS$ basis.
\item[4.] Express the $\tilde \psi^{\text {SOCI}}$ wave functions from the $\LMLSMS$ to the $\JMJ$ basis using a transformation matrix based on Clebsch-Gordan coefficients.
\item[5.] Finally, build the des Cloizeaux effective Hamiltonian \cite{Cloizeaux:1960a} in the $\JMJ$ basis.
\end{itemize} 

\subsection{\label{subsec:wavefunctions} Analysis of the effective wave functions\protect\\ }

The effective wave functions match the projected and orthogonalized \textit{ab initio} wave functions, by construction. Diagonalization of the SR effective Hamiltonian in the $\LML$ basis and of the SOCI effective Hamiltonian in the $\JMJ$ basis generates effective wave functions expressed in the corresponding $\LML$ and $\JMJ$ bases. \texttt{NewMag} prints both these effective wave functions and their corresponding energies. The latter must exactly match the \textit{ab initio} energies (by construction). Therefore, the user can quickly verify that the effective Hamiltonian construction has not been compromised. Also, since the compositions are computed in the $\JMJ$ basis, the weights associated with each individual $\JMJ$ function can be determined, allowing the $\left|J,\pm M_J\right>$ weights (which are printed by default by \texttt{NewMag}) to be summed, as well as the weights over all $\JMJ$ functions belonging to a given $J$ manifold. As exemplified in Section \ref{subsec:fluo_Ce}, this can be useful for understanding optical transitions.

\subsection{\label{subsec:extraction} Parameter extractions\protect\\ }

The extraction of the CF $B_k^q$ parameters at the SR level and of the $B_k^q$ and $\zeta$ parameters at the SOCI level follows the same procedure as described earlier \cite{NewMag_Sergentu:2026}, based on the irreducible tensor operator (ITO) procedure of Chibotaru and Ungur \cite{SingleAniso_Chibotaru:2012a, SingleAniso_Ungur:2017a}. At the SR level, the $B_k^q$'s are directly extracted from the effective Hamiltonian:

\begin{equation}
B_k^q (\text{SR}) = \frac{\text{Tr}(\mathbf{H^\text{eff}}(\text{SR})\cdot \mathbf{O_k^q})}{\text{Tr}(\mathbf{O_k^q} \cdot \mathbf{O_k^q})}
\end{equation}

\noindent The highest possible value for a $k$ rank relates to the degrees of freedom that arise from the size of the model space, comprising 2$L$+1 elements: it is simply equal to 2$L$ (2$L$ operations are necessary to couple the $M_L^\text{min}$ and $M_L^\text{max}$ configurations). The $k$ ranks are even, ranging from 2 to its maximum value, and $q$ orders range from $-k$ to $+k$ for a given $k$. Therefore, the total number of CF parameters can be easily derived from knowledge of the maximum value of $k$. All these data are summarized in Table \ref{ranks} for all the configurations of interest. Note that at the SR level, the number of CFPs is the same as with \texttt{SINGLE\_ANISO}. In practice, \texttt{NewMag} extracts the 90 CFPs for all the $f^n$ configurations of interest ($f^1$--$f^6$ and $f^8$--$f^{13}$), and the user may verify that the irrelevant parameters are strictly numerical zeros, if that must be the case. 

\begin{table}[h!]
\caption{The highest possible rank $k$ at the SR level and the resulting number of CFPs. Note that it also applies to the SOCI level with \texttt{NewMag}.\\}
\label{ranks}
\centering
\begin{tabular}{|c|cccccc|cccccc|}
\hline
$f^n$ & $f^1$  & $f^2$  & $f^3$  & $f^4$  & $f^5$  & $f^6$  & $f^8$  & $f^9$  & $f^{10}$  & $f^{11}$  & $f^{12}$  & $f^{13}$  \\
\hline
Highest $k$ & 6 & 10 & 12 & 12 & 10 & 6 & 6 & 10 & 12 & 12 & 10 & 6 \\
\hline
Number of CFPs & 27 & 65 & 90 & 90 & 65 & 27 & 27 & 65 & 90 & 90 & 65 & 27 \\
\hline
\end{tabular}
\end{table}

At the SOCI level, two operations are performed to extract parameters:

\begin{itemize}
\item[a.] The CF $B_k^q$ parameters are extracted from the SOCI effective Hamiltonian similar to the SR level:
\begin{equation}
B_k^q (\text{SOCI}) = \frac{\text{Tr}(\mathbf{H^\text{eff}}(\text{SOCI})\cdot \mathbf{O_k^q})}{\text{Tr}(\mathbf{O_k^q} \cdot \mathbf{O_k^q})}
\end{equation}
\item[b.] The ITO procedure is also applied to extract $\lambda (\text{SOCI})$:
\begin{equation}
\lambda (\text{SOCI}) = \frac{\text{Tr}(\mathbf{H^\text{eff}}(\text{SOCI})\cdot {(\mathbf L\cdot}\mathbf S))}{\text{Tr}({(\mathbf L\cdot}\mathbf S) \cdot {(\mathbf L\cdot}\mathbf S))}
\end{equation}
\end{itemize}

\noindent The ``monoelectronic'' SOC constant, positive defined, is finally obtained as follows:

\begin{equation}
\zeta (\text{SOCI}) = \pm 2S \lambda (\text{SOCI}) = \pm \tilde{n} \lambda (\text{SOCI})
\end{equation}

\noindent where the plus sign applies to less than half filled shells (\textit{i.e.} $f^1$--$f^6$) and the minus sign applies to more than half-filled shells (\textit{i.e.} $f^8$--$f^{13}$). 

The resulting SOC constants are similar to those obtained by \texttt{AILFT} since \texttt{NewMag} also uses the spherical approximation. Note that since AILFT only builds a CF Hamiltonian in the one-electron basis (which is actually printed in the basis of RSHs), the highest possible rank $k$ is 6, meaning that the model lacks some physics in many configurations, apart from $f^1$, $f^6$, $f^8$ and $f^{13}$. 

Comparing \texttt{NewMag} with \texttt{SINGLE\_ANISO}, which makes use of the pseudo-$J$ approximation, there are differences concerning the maximum value for $k$, if $k$ is defined directly from the size of the model space, \textit{i.e.} 2$J$+1. These values, as computed by \texttt{SINGLE\_ANISO}, are reported in Table \ref{ranksANISO}. Comparison with \texttt{NewMag} data in Table \ref{ranks} concludes that relevant parameters are missed by \texttt{SINGLE\_ANISO} in the $f^1$--$f^6$ cases, and too many parameters are introduced in the $f^8$--$f^{12}$ (only the f$^{13}$ configuration is consistent between the two codes). 

\begin{table}[h!]
\caption{The highest possible rank $k$ at the SOCI level within the pseudo-$J$ approximation, as done by \texttt{SINGLE\_ANISO}, and the resulting number of CFPs.\\}
\label{ranksANISO}
\centering
\begin{tabular}{|c|cccccc|cccccc|}
\hline
$f^n$ & $f^1$  & $f^2$  & $f^3$  & $f^4$  & $f^5$  & $f^6$  & $f^8$  & $f^9$  & $f^{10}$  & $f^{11}$ & $f^{12}$ & $f^{13}$  \\
\hline
Highest $k$ & 4 & 8 & 8 & 8 & 4 & 0 & 12 & 14 & 16 & 14 & 12 & 6 \\
\hline
Number of CFPs & 14 & 44 & 44 & 44 & 14 & 0 & 90 & 119 & 152 & 119 & 90 & 27 \\
\hline
\end{tabular}
\end{table}

This is completely bypassed by the \texttt{NewMag} approach, since the direct ITO extraction of the CFPs in the $\JMJ$ basis, based on the full manifold that correlates with the $^{2S+1}L$ ground state of the reference free ion, allows the extraction of the same number of CFPs as at the SR level (see Table \ref{ranks}). This aspect is perfectly consistent with the CF theory, since the CF should only act on the orbital degrees of freedom and since the SOC is treated as a perturbation of the SR picture within the SOCI framework. Key examples will be given in Section \ref{subsec:fn_case}, notably concerning the $f^6$ and $f^9$ configurations. Moreover, the \texttt{NewMag} approach naturally introduces the $J$-mixing \cite{NewMag_Sergentu:2026}, \textit{i.e.} mixing between components of different $J$ manifolds due to second-order CF couplings \cite{Abragam:1970a}. Such couplings are obtained here by construction of the CF Hamiltonian, which is built in the $\LMLSMS$ basis prior to transforming it to the $\JMJ$ one (full space, \textit{i.e.} no pseudo-$J$).

\subsection{\label{subsec:reconstruct} Reconstruction of the model spectrum and model quality\protect\\ }

It is instructive to reconstruct the model Hamiltonian based on all the extracted parameters at the SOCI level:

\begin{equation}
\label{re}
\mathbf{H^\text{mod}} = \mathbf{H^\text{mod,CF}} + \mathbf{H^\text{mod,SOC}}
\end{equation}

\noindent Diagonalization of this model Hamiltonian yields the model energies, which do not exactly match the \textit{ab initio} ones as a consequence of the approximate nature of $\mathbf{H^\text{mod,SOC}}$ (spherical approximation) \cite{NewMag_Sergentu:2026}. The mean absolute error (MAE) on the (2$L$+1)($\tilde n$+1) energy levels is then calculated and printed:

\begin{equation}
\text{MAE} = \frac {1}{(2L+1)(\tilde n+1)} \sum\limits_{j'} |\text E_{j'} ^\text{mod}-E_{j'} ^\text{SOCI}|
\end{equation}

\noindent where the $j'$ index is used instead of $j$ in Equation \ref{eq:SOCI} since the model space may span less states than the SOCI space. The error is further calculated as a percentage of the spectral width \cite{Bastardis:2007}, $\Delta E$ = $E_\text{max} ^\text{SOCI} - E_\text{min} ^\text{SOCI}$, as follows:

\begin{equation}
\text{MAEER} (\%) = \frac{\text{MAE}}{\Delta E} \times 100
\end{equation}

\noindent \texttt{NewMag} also prints the root mean square deviation (RMSD) to the \emph{ab initio} energies, defined as follows:

\begin{equation}
\text{RMSD} = \sqrt{\frac {\sum\limits_{j'} (\text E_{j'} ^\text{mod}-E_{j'} ^\text{SOCI})^2}{(2L+1)(\tilde n+1)}} 
\end{equation}

Finally, the impact of the various $k$ ranks on the model spectum is calculated, by reconstructing the model Hamiltonian with the SOC constant plus only $k$=2 parameters ($\lambda$ + 5 CFPs), $k$=2 + $k$=4 ($\lambda$ + 14 CFPs), $k=$2 + $k$=4 + $k$=6 ($\lambda$ + 27 CFPs), and so on, up to $k$=2 + ... + $k$=12 ($\lambda$ + 90 CFPs). The process is repeated at the SR level, based on the 2$L$+1 $i'$ SR roots (also note the $i'$ notation instead of the $i$ one in Equation \ref{eq:CAS}), in the $\LML$ basis and absence of the SOC operator (that is, only with $\mathbf{H^\text{mod}}= \mathbf{H^\text{mod,CF}}$).

\subsection{\label{subsec:misc} Miscellaneous: Hints and tricks\protect\\ }

Most of the \texttt{NewMag} machinery has already been exposed. However, a few subtleties still deserve to be exposed, especially for advanced users or readers who are interested in implementing a similar approach. 

First of all, there must be a correspondence between the phases of the SR states and the conventions that are applied within all the transformation matrices of interest (in particular the $U$ and $C$ matrices of Section \ref{subsec:SR_H}). In \texttt{NewMag}, $U$ and $C$ were hard-coded with the Condon-Shortley convention and \textit{ad hoc} corrections of the $\tilde \psi^{\text {SR}}$ wave functions have been implemented. Note that phases are a known issue when dealing with effective off-diagonal elements \cite{Gould, Sergentu:2024, NewMag_Sergentu:2026, Suaud:2026}.

As mentioned by others \cite{Duros:2025a}, CFPs are meaningless if the used convention and normalization is not specified. We have used non-normalized ``extended'' Stevens operators, and applied $\left< L|| \alpha ||L\right>$,  $\left< L|| \beta ||L\right>$ and  $\left< L|| \gamma ||L\right>$ prefactors\cite{Abragam:1970a} for rank $k$ = 2, 4 and 6 operators, respectively. Higher-rank terms were directly extracted with no prefactor, as it is done with \texttt{SINGLE\_ANISO} at the SR level. We have not applied $\left< J|| \alpha ||J\right>$,  $\left< J|| \beta ||J\right>$ and $\left< J|| \gamma ||J\right>$ prefactors at the SOCI level since by construction we have applied the  $\left< L|| \alpha ||L\right>$,  $\left< L|| \beta ||L\right>$ and $\left< L|| \gamma ||L\right>$  ones prior to explicitly transforming the model Hamiltonian into the $\JMJ$ basis, meaning that within given $J$ blocks the results are consistent with the direct application of the $\left< J|| \alpha ||J\right>$,  $\left< J|| \beta ||J\right>$ and $\left< J|| \gamma ||J\right>$ prefactors with the Stevens operators acting on the $\JMJ$ functions. Therefore, the \texttt{NewMag} data is directly comparable with data from \texttt{SINGLE\_ANISO}, at both the SR and SOCI levels. 

To establish comparison with \texttt{AILFT}, additional capability was implemented in \texttt{NewMag}. At the SR level, \texttt{AILFT} extract parameters from the CI matrix, mapping monoelecronic integrals into a monoelectronic CF model and bielectronic electron-electron integrals on simple models such as Racah's one. For the part that we are interested in, the CF Hamiltonian is printed in the output of \texttt{AILFT} in the basis of the RSHs within a monoelectronic picture. We use this 7$\times$7 matrix to extract 27 CFPs (up to rank-6 parameters) as follows:

\begin{itemize}
\item[1.] The \texttt{AILFT} CF Hamiltonian is transformed from the RSHs to the CSHs, simply using $U_1$ (recall, due to the monoelectronic picture, diagonalization of the CF Hamiltonian leads to 7 one-electron energies, \textit{i.e.} the energies of the $f$ orbitals).
\item[2.] The 27 CFPs are extracted with the present ITO procedure. Here, the $\left< l|| \alpha ||l\right>$,  $\left< l|| \beta ||l\right>$ and  $\left< l|| \gamma ||l\right>$ prefactors are incorporated to ensure consistency of the SR and SOCI parameters produced by \texttt{NewMag}.
\end{itemize}

\noindent Overall, \texttt{NewMag} can be used to compare results obtained with \texttt{AILFT}, \texttt{NewMag} and \texttt{SINGLE\_ANISO}, since the CFPs are all consistently defined for comparison purposes, with no phase, convention or prefactor divergence.

Several aspects may render the CFPs difficult to rationalize:

\begin{itemize}
\item[a.] The number of CFPs can be large (see Table \ref{ranks}).
\item[b.] Each individual CFP can be altered by a change of the coordinate frame.
\item[c.] There is no ``natural'' or ``good'' frame for an asymmetric system.
\end{itemize}

\noindent Therefore, an asset is to define rotationally invariant parameters \cite{Chang, Leavitt} by considering ``normalized'' Stevens parameters. Therefore, using all the rotationally invariant parameters defined below, and their components, the extracted CFPs are re-scaled to ensure normalization, following Rudowicz \cite{Rudowicz:1985}. Since the normalization factors were only available up to rank-6 terms, we have only used the rank-2, rank-4 and rank-6 parameters to define rotationally invariant parameters. Note that the resulting $B_k^{\prime q}$ parameters are not printed by \texttt{NewMag}, even if they are used to compute the $S$ indicator, which is a measure of the crystal field ``strength'', as well as its constituents. $S$ can be derived in two ways \cite{Jung}:

\begin{itemize}
\item[a.] From the rank-by-rank contributions:

\begin{equation}
S = \sqrt{\frac {(S_2)^2 + (S_4)^2 + (S_6)^2}{3}} =  \sqrt{\frac {\sum\limits_{k=2}^6 (S^k)^2}{3}} 
\end{equation}

\item[b.] From the order-by-order contributions:

\begin{equation}
\label{o}
S =  \sqrt{\frac {\sum\limits_{q=0}^6 (S_q)^2}{3}} 
\end{equation}

\end{itemize}

\noindent Both definitions are strictly equal, provided that the Stevens parameters have been properly ``normalized''. Note that the expressions of the $S_k$ and $S^q$ parameters are given elsewhere \cite{Jung}. While $S_2$, $S_4$ and $S_6$ are also rotationally invariant, it is not generally the case for the $S^0$, $S^1$, ..., and $S^6$ parameters \cite{Jung}. This will be exemplified in Section \ref{subsec:Eu-Zn}. 

\section{\label{sec:result-discuss}Computational details \protect\\ }

In this article, we report six case studies aiming at illustrating the capabilities of the current \texttt{NewMag} development for lanthanide complexes (four cases), as well as exploring potential limitations when applying the code for other purposes, for instance to study transition metal complexes (two cases). Because the implementation works with both OpenMolcas and ORCA, results generated with both these programs are reported, but a detailed comparison between data generated with these two programs is given only in Section \ref{subsec:Dy-Cp2}. The systems are presented later; therefore, this section only provides the generic computational setups, with system-specific details given in Section~\ref{sec:result-discuss}. All calculations were performed without symmetry, \textit{i.e.}, using the $C_1$ point group regardless of the actual symmetry of the system. The output files analyzed in this article are available at \url{https://github.com/clausserg/newmag}, where further details on the individual calculations can be found.

\subsection{\label{subsec:MOLCAS}OpenMolcas calculations \protect\\ }

All calculations performed with OpenMolcas\cite{OpenMolcas_Manni:2023a} used the release version 25.06. The Douglas-Kroll-Hess Hamiltonian\cite{Douglas:1974, Hess:1986, Jansen:1989} was employed, together with the ``compatible with it'' ANO-RCC basis sets\cite{ANORCC-1, ANORCC-2, ANORCC-3}. SA-CASSCF calculations\cite{CASSCF_Roos:1980a,CASSCF_Roos:1980b} were typically performed with ``same-spin'' SR roots, unless specified otherwise. If two sets of spin states are considered at the SR level, note that OpenMolcas build them with two sets of SA orbitals, unlike ORCA. Dynamically-correlated energies were obtained with CASPT2 \cite{ss_CASPT2_Andersson:1990a,ss_CASPT2_Andersson:1992a}. An imaginary shift\cite{Forsberg:1997} of 0.2 a.u. was employed to prevent the occurrence of intruder states, while the IPEA shift\cite{Ghigo:2004} was set to 0. The SOC was computed based on the atomic mean-field integrals (AMFI)\cite{Hess:1996} with the RASSI-SO method\cite{RASSI_Malmqvist:2002a,Roos:2004}. Note that by default, 2$L$+1 SR roots were in fact computed, generating $\sum\limits_J$2$J$+1=(2$L$+1)($\tilde n$+1) SOCI roots (see Table \ref{roots}). When applicable, \texttt{SINGLE\_ANISO}\cite{SingleAniso_Chibotaru:2012a} was called for, though we have not reported those results in this article since similar conclusions are obtained with ORCA. 

\subsection{\label{subsec:ORCA}ORCA calculations \protect\\ }

All calculations performed with ORCA\cite{Orca_Neese:2025a} used version 6.1.0. The Douglas-Kroll-Hess Hamiltonian\cite{Douglas:1974, Hess:1986, Jansen:1989} was also employed, together with appropriate basis sets: DKH-def2 basis sets \cite{Weigend:2005a, Pantazis-1} for light (H, C, N, O) as well as for transition metal atoms (Zn) and SARC2-DKH basis sets\cite{Aravena:2016b} for the lanthanide atoms (Eu, Dy). SA-CASSCF calculations\cite{CASSCF_Roos:1980a,CASSCF_Roos:1980b} were performed prior to performing SC-NEVPT2 ones\cite{NEVPT2_Angeli:2001a} to generate dynamically-correlated energies. As in the OpenMolcas calculations, 2$L$+1 SR roots were usually computed. The SOC was computed based on a mean-field approximation\cite{Neese:2005a} with a standard SOCI scheme (\textit{i.e.} diagonalization). Both \texttt{AILFT}\cite{ailft_Atanasov:2015a} and \texttt{SINGLE\_ANISO}\cite{SingleAniso_Chibotaru:2012a} calculations were performed.

\section{\label{sec:result-discuss}Results and discussion \protect\\ }
\subsection{\label{subsec:f1_case} Back to the $\bm{f^1}$ case \protect\\ }

\subsubsection{\label{subsec:urano_cerno} Impact of electron correlation on the CFPs of cerocene  \protect\\ }
\begin{figure}
    \centering
    \includegraphics[width=0.75\linewidth]{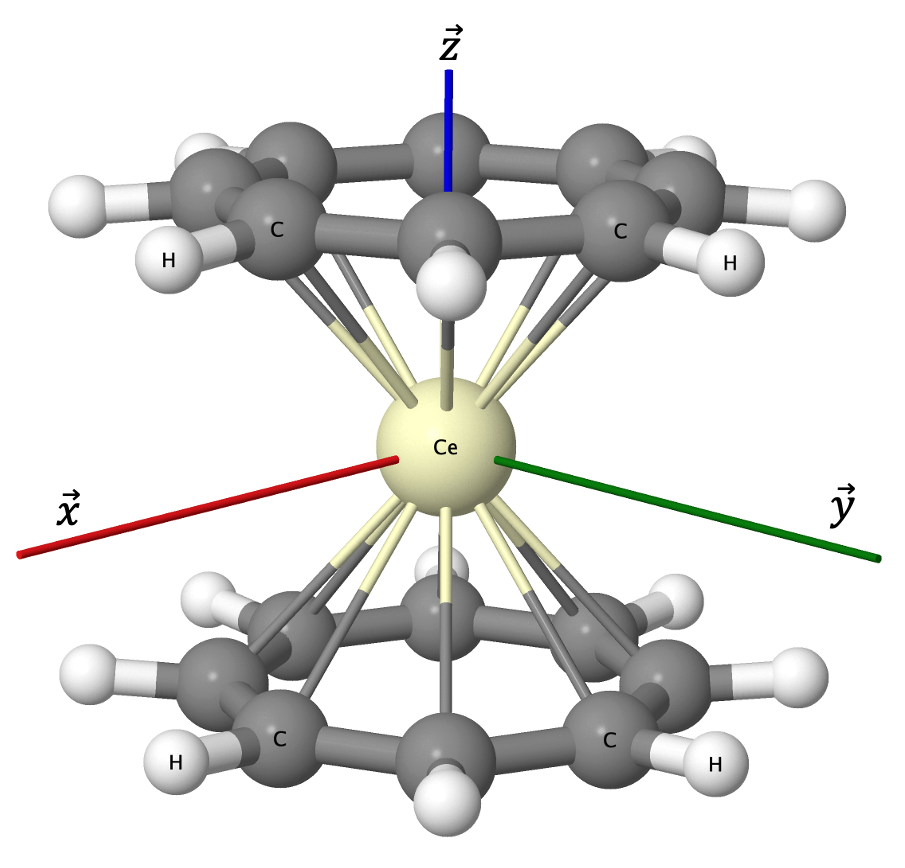}
    \caption{Representation of the $D_{8h}$ structure of $\left[ \text{Ce}(\text{C}_{8}\text{H}_{8})_{2} \right]^{-}$. The $\vec{z}$ axis is aligned with the $C_8$ symmetry axis ($\vec{x}$ and $\vec{y}$ are arbitrary in this symmetry point group). Color code: Ce = yellow, C = gray and H = white.}
    \label{cero}
\end{figure}

\indent In this part, the impact of electron correlation on the CFPs is assessed. To this end, we revisit the cerocene anion, \textit{i.e.} the bis(cyclooctatetraenyl)cerium(III) anion, $\left[ \text{Ce}(\text{C}_{8}\text{H}_{8})_{2} \right]^{-}$ \cite{Walter:2009a}. This system is a playground to modeling the crystal field potential within the theoretical community \cite{Le_Roy:2014,Gendron:2015,Gendron:2019} and was already included in our first paper\cite{NewMag_Sergentu:2026} on \texttt{NewMag}. It displays a $D_{8h}$ structure (see Figure \ref{cero}). Because of the $f^1$ configuration, only ranks 2, 4 and 6 are allowed (see Table \ref{ranks}). In this case, the $D_{8h}$ symmetry point group leads to pure axiality (only the $k$=0 is allowed), leading to three symmetry-allowed parameters: $B_2^0$, $B_4^0$ and $B_6^0$. 

The $^2F$ term of the Ce(III) free-ion splits into $^2F_{5/2}$ and $^2F_{7/2}$ due to SOC, with $^2F_{5/2}$ being the ground energy level. In the complex, in the presence of the axial (anisotropic) CF, the two terms mix \cite{Gendron:2015,Gendron:2019,Cespedes:2017a}. With \texttt{NewMag}, we have shown that such $J$-mixing, due to second-order CF couplings, can be fully captured by constructing the model in the $\LMLSMS$ basis prior to transforming it in the $\JMJ$ one\cite{NewMag_Sergentu:2026}. In the literature, an extraction omitting the $B_6^0$ parameter was performed before \cite{Cespedes:2017a}, which considered the same highest $k$ as what would be done within pseudo-$J$ approximation (see Table \ref{ranksANISO}). Our previous extraction was then not only the first one that included $B_6^0$, but also the first one with an explicit and consistent handling of the $J$-mixing. However, we considered only minimal active space calculations and only SA-CASSCF and SO-CASSCF calculations\cite{NewMag_Sergentu:2026} because of limitations of the first released version. Here, we aim at considering a larger active space and/or more correlated energies based on second-order perturbation theory, with OpenMolcas, and with triple-$\zeta$ quality basis sets on all atoms (ANO-RCC-VTZP). 

Two active spaces were considered. The first one is the minimal active space, comprising 1 electron within 7 orbitals, denoted CAS(1,7). The active orbitals are essentially of $4f$ character. The second one, denoted CAS(5,9), augments the minimal CAS with the bonding orbitals between the ligand $\pi_\delta$ orbitals and the lanthanide $4f_{\delta}$ ones, denoted $(\pi_\delta$--$4f_{\delta})_+$ where the ``+'' symbol indicates in-phase bonding character\cite{Sergentu:2018}. Of course, the bonding orbitals are occupied in the ground configuration, meaning that their inclusion adds 4 electrons and 2 orbitals to the active space on top of the minimal CAS. With a ``balanced'' extension of the active space, it is expected to introduce additional correlation already at the SA-CASSCF level (complemented by further CASPT2 calculations). The rationale behind such extension of the active space is to balance the description of covalency involving the metal $4f_{\delta}$ AOs, which may affect both the CFPs and the SOC constant. Ideally, the results should converge with respect to the size of the active space. \\
\begin{table}[h!]
\caption{Parameter values in cm$^{-1}$ for the $\left[ \text{Ce}(\text{C}_{8}\text{H}_{8})_{2} \right]^{-}$ complex, with the $\left|L,M_L\right>$ or the $\left|J,M_{J}\right>$ basis (full space), as a function of the active space. CFPs with values below 0.1 cm$^{-1}$ in the \texttt{NewMag} outputs are skipped.\\}
\label{CFPs_cerocene_NewMag}
\centering
\begin{tabular}{|c|cccc|}
\hline
\multicolumn{5}{|c|}{CAS(1,7)}            \\         
\hline
   & SR-CASSCF & SO-CASSCF & SR-CASPT2 & SO-CASPT2 \\
\hline
$B_2^0$ & $-$336.9   & $-$336.9   & $-$467.0   & $-$467.0   \\
$B_4^0$ & $-$697.3   & $-$697.3   & $-$745.4   & $-$745.4   \\
$B_6^0$ & 54.7    & 54.7     & 58.8     & 58.8     \\[0.1cm] 
$\zeta{(4f)}$& n/a  & 671.1    & n/a       &671.1    \\[0.1cm] 
MAE & 0.0 & 4.6 & 0.0 & 4.6 \\[0.1cm] 
\hline
\multicolumn{5}{|c|}{CAS(5,9)}              \\         
\hline
    & SR-CASSCF & SO-CASSCF & SR-CASPT2 & SO-CASPT2 \\
\hline
$B_2^0$ & $-$335.8   & $-$335.8  & $-$458.7   & $-$458.7   \\
$B_4^0$ & $-$700.9   & $-$700.9  & $-$774.3  & $-$774.3  \\
$B_6^0$ & 55.5     & 55.5    & 57.6    & 57.6    \\[0.1cm]
$\zeta{(4f)}$& n/a  & 671.0    & n/a       &671.0    \\[0.1cm] 
MAE & 0.0 & 4.3 & 0.0 & 4.3 \\[0.1cm] 
\hline 
\end{tabular}
\end{table}

The SR-CASSCF, SO-CASSCF, SR-CASPT2 and SO-CASPT2 results obtained with the two active spaces are reported in Table \ref{CFPs_cerocene_NewMag}. As already mentioned, only 3 CFPs are allowed in the $D_{8h}$ symmetry point group with the $f^1$ configuration. However, since \texttt{NewMag} computes anyway the maximum number of CFPs that can occur for the $f^1$--$f^6$ and $f^8$-$f^{13}$ configurations (90, see Table \ref{ranks}), and since the \textit{ab initio} calculations are performed in the $C_1$ symmetry point group, some of the remaining 87 CFPs could be numerically non-zero. In practice, very small values were obtained for any of these spurious terms, all being well below 0.1 cm$^{-1}$, that is several orders of magnitude smaller than the reported terms. Hence, those terms are not reported, but can be found in the outputs (see the \texttt{NewMag} repository on GitHub \cite{NewMag:version1}). Although these parameters have little impact, all such near-zero parameters are nevertheless included in the reconstruction of the model spectra. One could naively think that it would be wiser to consider symmetry in the calculations. However, high-symmetry point groups such as $D_{8h}$ are not implemented in standard quantum chemistry codes and using a subgroup of a high-symmetry point group may sometimes lead to undesired behavior because degenerate orbitals may appear in distinct irreducible representations. Therefore, we chose to consider the $C_1$ symmetry point group for all calculations reported here, and thus symmetry is not explicitly handled by \texttt{NewMag}. To be complete in the description of \texttt{NewMag}, the zero of the energy was fixed at the lowest SR state or at the lowest energy level when reconstructing the model spectra, which for non-perfect reproductions, may affect the error committed, \textit{e.g.} the MAE in Table \ref{CFPs_cerocene_NewMag} and in other Tables as well. 

Let us start by analyzing the SR-CASSCF and SO-CASSCF data obtained with CAS(1,7). With SA-CASSCF, 7 SR roots are computed, which converts into 14 spin-orbit roots. At the SOCI level, these roots occur in Kramers doublet pairs, meaning that 7 distinct \textit{ab initio} energies are computed. Though the numbers differ from our previous paper \cite{NewMag_Sergentu:2026}, essentially because of basis set effects, we also extract the exact same CFPs at the SR-CASSCF and SO-CASSCF levels. In fact, this is something that we also observe at the SR-CASPT2 and SO-CASPT2 levels, and also with the larger active space, CAS(5,9). This is again a positive signal for the present development, since it should not be otherwise here since the SOC is introduced as a perturbation of the SR Hamiltonian and since the model and the SOCI spaces perfectly match in size and essentially in nature, meaning that our model should not miss any relevant SOC. 

Regardless of the SR level, the model Hamiltonian perfectly reproduces the \textit{ab initio} energies as soon as ranks 2, 4 and 6 are included (MAE = 0.0 cm$^{-1}$). This is not the case with SOCI, where deviations are fully attributed to the SOC being modeled within the spherical approximation and assuming pure $4f$ orbitals (CF picture). However, MAE values of about 4.5 cm$^{-1}$ are obtained, while the spectral width, $\Delta E$ = $E_\text{max} ^\text{SOCI} - E_\text{min} ^\text{SOCI}$, is $\sim$4300 cm$^{-1}$. In other words, our simple SOC model Hamiltonian is quantitative.

Since the CFPs are the same at the same SR and SOCI levels, the role of the CASPT2 correlation can be discussed directly by comparing the SO-CASSCF and SO-CASPT2 data. At the SO-CASPT2 level, CASPT2 correlation only modifies the diagonal elements of the SOCI matrix, while the off-diagonal elements are still computed from the SA-CASSCF wave functions. Therefore, it is expected that CASPT2 has no effect on the SOC constants. The CFPs, however, behave differently: because CASPT2 modifies the energies of the SR roots, the resulting CFPs are also necessarily revised. Since the signs and orders of magnitude of the individual CFPs are maintained, CASPT2 here improves the quantitative picture, without revising it qualitatively. It is clear from Table~\ref{CFPs_cerocene_NewMag} that the two active spaces yield very similar results, regardless of the level of theory. This can be attributed to the predominantly core-like character of the $4f$ orbitals and, consequently, to the generally weak covalent character of lanthanide bonds. \\

A last interesting point concerns the $J$-mixing. Since the system is axial, only same-$M_J$ components are expected to mix, meaning that the $\left|7/2,\pm7/2\right>$ components (and only those) should remain pure (\textit{i.e.} be $J$-mixing free). This behavior is reproduced by \texttt{NewMag} regardless of the employed level of theory. Other levels mix two-by-two. Since the $J$-mixing remains moderate, each Kramers doublet has a net dominant $\left|5/2\right>$ or $\left|7/2\right>$ character. In fact, the multiplet structure of the reference free ion is maintained, the first 3 Kramers doublet are dominated by the $\left|5/2\right>$ components while the following 4 doublets are dominated by the $\left|7/2\right>$ character. Therefore, a ``blind'' pseudo-$J$ approximation would have selected the correct roots, which is not surprising for a lanthanide complex, where SOC is much stronger than the CF, consistent with the predominantly core-like character of the $4f$ orbitals. For the ground energy level, CASPT2 slightly enhances the $J$-mixing, from 1.3\% at the SO-CASSCF levels to 1.5--1.8\% at the SO-CASPT2 levels. This can be rationalized from the data in Table~\ref{CFPs_cerocene_NewMag}: all CFPs increase in absolute value, while the SOC constant remains unchanged. Thus, the second-order CF couplings between the $\left|5/2\right>$ and $\left|7/2\right>$ blocks are enhanced, whereas the difference between the means of their diagonal elements remains unchanged. Consequently, $J$-mixing is enhanced for all energy levels, not only for the ground energy level. The $J$-mixing is also more pronounced for the excited states, reaching, for example, $\sim$15\% for the third Kramers doublet. \\

\noindent \fbox{%
\begin{minipage}{0.46\textwidth}
Overall, we have successfully reported an implementation of \texttt{NewMag} that is operative with OpenMolcas, extended active spaces, and with CASPT2 and SO-CASPT2. In $\left[ \text{Ce}(\text{C}_{8}\text{H}_{8})_{2} \right]^{-}$, the minimal active space captures the essential physics, while CASPT2 adds a significant improvement of the CFPs. Since similar results are expected in most if not all the lanthanide complexes, only minimal active space calculations are retained in the remainder of the article. Although this conclusion may seem somewhat unsurprising, the SO-CASSCF approach already provides a good qualitative picture of lanthanide complexes. It may therefore remain a useful alternative when CASPT2 calculations are computationally too demanding, as can be the case for larger systems.
\end{minipage}
}

\subsubsection{\label{subsec:fluo_Ce}  Unraveling the fluorescence of the Ce$^\text{III}$-aqua complex \protect\\ }

The second case study aims at illustrating how \texttt{NewMag} can help in rationalizing luminescence properties of lanthanide complexes. We have retained the case of the cerium(III)-aqua complex, previously reported by Lindqvist-Reis \textit{et al.} \cite{Lindqvist-Rei:2018}. In the ground state, this system displays nine water molecules in the first coordination sphere of the Ce$^\text{III}$ ion. After photoexcitation, one of the coordinated waters moves to the second coordination sphere, leading to the $\left[ \text{Ce}^*(\text{H}_{2}\text{O})_{8}\cdot (\text{H}_{2}\text{O})\right]^{3+}$ complex. Given the energy involved, a 4$f^1$ $\rightarrow$ 5$d^1$ electronic transition was expected, which was confirmed by quantum mechanical calculations \cite{Lindqvist-Rei:2018}. In the computed emission spectrum, two bands constitutive of the broadened transition were described as decay to $^2F_{7/2}$ and to $^2F_{5/2}$. In this article, discrete CF levels were not resolved. However, if one assumes that the emissive state correlates with the Ce(III) $^2D_{3/2}$, only one band should be active ($\Delta J=0,\pm1$). \texttt{NewMag} is used here to shed light on this aspect, in particular by determining the state compositions in terms of $^2F_{5/2}$/$^2F_{7/2}$ and $^2D_{3/2}$/ $^2D_{5/2}$ characters, \textit{i.e.} by estimating the $J$-mixing in the components of both the manifolds. Note that a recent preprint, made available after the present work, reports an alternative to \texttt{NewMag} to estimate the $J$-mixing\cite{preprint}.

\begin{figure}
    \centering
    \includegraphics[width=0.75\linewidth]{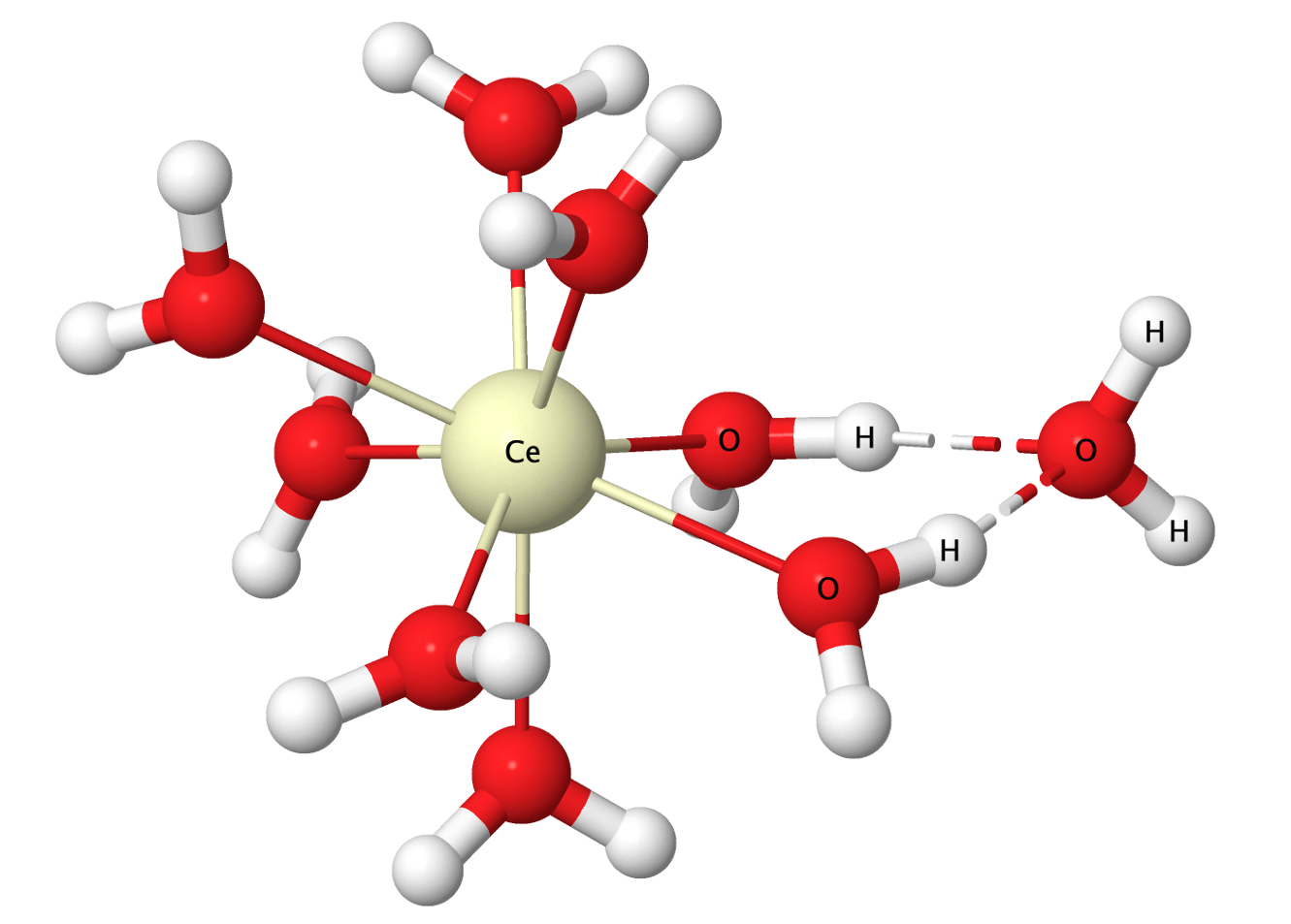}
    \caption{Representation of the $C_{1}$ structure of $\left[ \text{Ce}^*(\text{H}_{2}\text{O})_{8}\cdot (\text{H}_{2}\text{O})\right]^{3+}$. Color code: Ce = yellow, O = red and H = white.}
    \label{aquo}
\end{figure}

First, we have determined a genuine structure for the $\left[ \text{Ce}^*(\text{H}_{2}\text{O})_{8}\cdot (\text{H}_{2}\text{O})\right]^{3+}$ complex, based on TD-DFT, obtained by optimizing the 8$^\text{th}$ SR root (the first 7$^\text{th}$ corresponding to the $4f^1$ manifold). The structure is displayed in Figure \ref{aquo} (coordinates are given in the GibHub repository of \texttt{NewMag}). Without going into detail, we employed a continuum solvent model~\cite{Barone:1997}, a standard exchange--correlation functional (PBE0~\cite{Adamo:1999}), and standard basis sets~\cite{Dolg:1989,Weigend:2005a}. The resulting structure is very similar to that reported by Lindqvist \textit{et al.}~\cite{Lindqvist-Rei:2018}. 

Next, SR-CASSCF, SO-CASSCF, SR-CASPT2 and SO-CASPT2 calculations were performed using the minimal active space correlating 1 electron in 12 orbitals, \textit{i.e.} CAS(1,12). As for the cerocene anion, ANO-RCC-VTZP basis sets were employed. To construct all states of the $4f^1$ and $5d^1$ manifolds, 12 SR roots were computed, yielding 24 SOCI roots. Owing to Kramers degeneracy, this corresponds to 12 distinct SOCI energies, with the SO-CASPT2 values reported in Table \ref{Ce_FLUOR_mixing}. For wave-function analysis, since \texttt{NewMag} is designed to treat one manifold at a time, two separate \texttt{NewMag} calculations were performed: one for the $4f^1$ manifold, including 7 SR roots and 14 SOCI roots, as in the case of the cerocene anion, and one for the $5d^1$ manifold, including 5 SR roots and 10 SOCI roots, namely the SR roots 8--12 and SOCI roots 15--24 in energetic order. As discussed later in Section \ref{subsec:dn_case}, the implementation is also directly applicable to the $5d^1$ manifold. Furthermore, no significant net $4f^1$/$5d^1$ mixing was observed at the SR level, with each root exhibiting more than 99.5\% projection onto the corresponding model space. The two manifolds can therefore be analyzed independently using \texttt{NewMag}. Since the complex is asymmetric, all 27 CFPs are active. These are not reported here, as the same conclusions as for the cerocene anion can be drawn, whereas the focus of this section is on the model wave functions.

\begin{table}[]
\caption{SO-CASPT2 energies (cm$^{1}$) and model compositions (\%) of the 12 computed energy levels (Kramers doublets) of $\left[ \text{Ce}^*(\text{H}_{2}\text{O})_{8}\cdot (\text{H}_{2}\text{O})\right]^{3+}$ derived by \texttt{NewMag} and their correlation with the $^2F$ and $^2D$ terms of the Ce$^{3+}$ free ion.\\}
\label{Ce_FLUOR_mixing}
\begin{tabular}{|c|c|c|}
\hline
$^{2S+1}L $  & $  E^\text{SOCI}  $ &Compositions\\
\hline
             & $\textcolor{red}{48442}$ &$  \textcolor{red}{5.4 \left|J=3/2\right>+ 94.6\left|J=5/2\right>}$ \\

              & $ \textcolor{red}{47467}$ & $\textcolor{red}{36.0 \left|J=3/2\right>+ 64.0\left|J=5/2\right>}$ \\

\textcolor{red}{$^{2}D$}                  & $ \textcolor{red}{46405}$ & $\textcolor{red}{21.0 \left|J=3/2\right>+ 79.0\left|J=5/2\right>}$ \\

              & $ \textcolor{red}{44353}$ & $\textcolor{red}{92.6\left|J=3/2\right>+ 7.4\left|J=5/2\right>}$ \\
 
Emissive                 & $ \textcolor{red}{24821}$ & $\textcolor{red}{45.0 \left|J=3/2\right>+ 54.0\left|J=5/2\right>}$ \\

\hline

            & $  \textcolor{blue}{3509}$ & $  \textcolor{blue}{0.2 \left|J=5/2\right>+99.8 \left|J=7/2\right>}$ \\

                  & $  \textcolor{blue}{3332}$ & $  \textcolor{blue}{0.2 \left|J=5/2\right>+99.8 \left|J=7/2\right>}$ \\

                & $  \textcolor{blue}{2727}$ & $ \textcolor{blue}{10.6 \left|J=5/2\right>+89.4\left|J=7/2\right>}$ \\

\textcolor{blue}{$^{2}F$}                  & $ \textcolor{blue}{2407}$ & $  \textcolor{blue}{1.3 \left|J=5/2\right>+98.7\left|J=7/2\right>}$ \\
\cline{2-3}
                & $ \textcolor{blue}{1064}$ & $\textcolor{blue}{99.6 \left|J=5/2\right>+0.4 \left|J=7/2\right>}$ \\
                 & $ \textcolor{blue}{605}$ & $ \textcolor{blue}{89.3 \left|J=5/2\right>+10.7\left|J=7/2\right>}$ \\

                & $ \textcolor{blue}{0}$ & $ \textcolor{blue}{98.8 \left|J=5/2\right>+1.2\left|J=7/2\right>}$ \\

\hline

\end{tabular}
\end{table}

From Table \ref{Ce_FLUOR_mixing}, it is clear that the lowest-energy three Kramers doublets are mainly of $\ket{J=5/2}$ character, followed by four doublets of main $\ket{J=7/2}$ parentage. Hence the horizontal line in the Table highlights the correlation with the two $\ket{J=5/2}$ and $\ket{J=7/2}$ terms of the reference free ion. The highest-energy five Kramers doublets correlate with $^2D$. The model compositions derived by \texttt{NewMag} reveal no clear structure of the reference free ion terms: it is practically impossible to define a set of two Kramers doublets correlating essentially with $\ket{J=3/2}$, and a set of three doublets correlating with $\ket{J=5/2}$. The emissive state is in fact composed almost equally of $\ket{J=3/2}$ and $\ket{J=5/2}$ character. As a consequence, it can, in principle, decay to any state of the $4f^1$ manifold, irrespective of whether that state correlates with $\ket{J=5/2}$ or $\ket{J=7/2}$ of the $^2F$ term. Moreover, such an approximately equal admixture intuitively suggests the possibility of two intense bands, as reported by Lindqvist-Reis \textit{et al.} \cite{Lindqvist-Rei:2018}, although the two bands would be expected to merge upon broadening.

\noindent \fbox{%
\begin{minipage}{0.46\textwidth}
Analysis of the \texttt{NewMag} model compositions of the energy levels makes it possible to readily resolve the discrete levels, including within excited-state manifolds. This provides a direct way to understand why certain transitions can be particularly intense in complexes, without relying much on their presumed correlation with the reference free-ion terms. Such an analysis highlights the continued relevance of CF theory and allows its full descriptive power to be exploited beyond standard, highly simplified treatments.
\end{minipage}
}

\subsection{\label{subsec:fn_case} Generalization to other $\bm{f^n}$ configurations \protect\\ }

For implementation testing, we considered at least one case for each of the $f^2$--$f^6$ and $f^8$--$f^{13}$ configurations. However, reporting all of these cases would be excessively lengthy; instead, we focus on two representative configurations: $4f^9$, which can typically lead to good SIMs, and $4f^6$, which is comparatively overlooked by \texttt{SINGLE\_ANISO} for obvious reasons (see Table \ref{ranksANISO}).

\subsubsection{\label{subsec:Dy-Cp2}Application to a model dysprosium(III) complex ($\bm{f^9}$)  \protect\\ }

Dysprosium(III) complexes belong to the $4f^9$ configuration. In this case, the \texttt{NewMag} workflow requires 11 SR roots (given the $^6H$ SR root of the reference free ion) which generates 66 energy levels at the SOCI level (correlating with $^6H_{15/2}$, $^6H_{13/2}$, ..., and $^6H_{5/2}$ of Dy$^{3+}$) and thus 33 distinct energies (Kramers degeneracy). In principle, $k$=10 can be reached (see Table \ref{ranks}). This is an ideal configuration for testing the \texttt{NewMag} development and comparing the results with those obtained using both \texttt{SINGLE\_ANISO} and \texttt{AILFT}. Results obtained with both OpenMolcas and ORCA are also presented in this section.

From the vast number of dysprosium(III) complexes reported in the literature, the $\left[\text{Dy}(\text{C}_{5}\text{H}_{5})_{2}\right]^{+}$ model complex was selected here because it was included in a previous extensive theoretical study\cite{Riccardo:2018a} and, among the systems investigated therein, it displayed two particularly relevant features: the largest splitting of the $^6H_{15/2}$ manifold, and hence the largest crystal-field ``strength'', as well as a ground $\ket{J=15/2, M_J=\pm15/2}$ Kramers doublet, making it a prototypical example of a ``good'' SIM.

\begin{figure}
    \centering
    \includegraphics[width=0.75\linewidth]{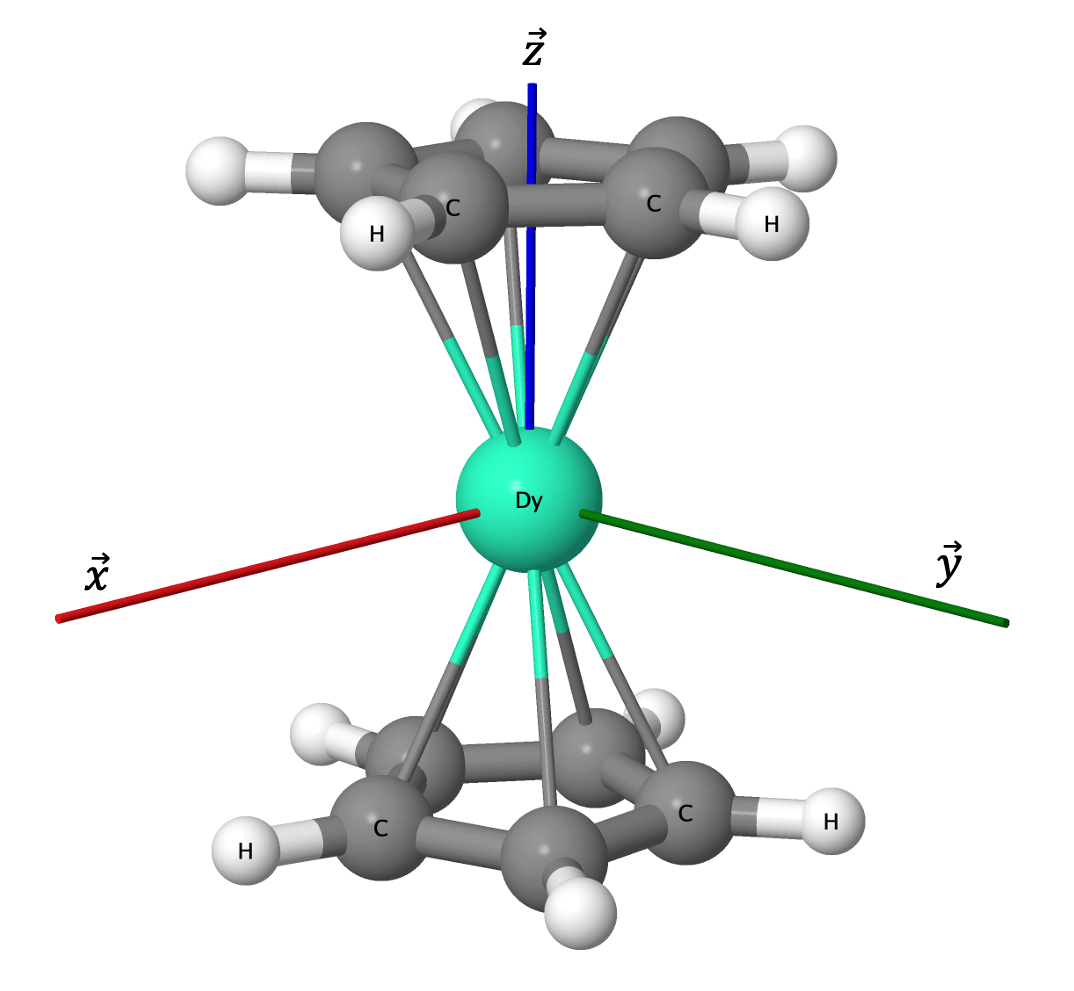}
    \caption{Representation of the $D_{5d}$ structure of $\left[ \text{Dy}(\text{C}_{5}\text{H}_{5})_{2} \right]^{+}$. The $\vec{z}$ axis is aligned with the $C_5$/$S_{10}$ symmetry axis, $\vec{x}$ is aligned with one $C_2$ axis and $\vec{y}$, constrained by the other two, is ion one $\sigma_d$ plane. Color code: Dy = green, C = gray and H = white.}
    \label{dyspro}
\end{figure}

Here, the structure reported by Alessandri \textit{et al.} was retained (see Figure \ref{dyspro}). It displays a $D_{5d}$ symmetry. In this symmetry, only $q$=0 and $|q|$=5 terms are allowed \cite{AllowedCF}. With the employed coordinate frame, the $B_k^5$ terms vanish and only the $B_k^{-5}$ terms are non-zero. Note that this can be inverted by inverting the $x$ and $y$ Cartesian axes. A $B_k^{-5}$ term can only exist if $k\ge6$, and as already mentioned, $k$ cannot be larger than 10 for an $f^9$ case. Therefore, in principle, 8 non-zero CFPs are expected: five axial $B_2^0$--$B_{10}^0$ CPFs and three $B_6^{-5}$--$B_{10}^{-5}$ CFPs. It turns out that only the terms of rank 2--6 are sizable, and thus, only 4 CFPs are displayed in Tables \ref{CFPs_dyspro_NewMag} and \ref{CFPs_dyspro_comp}.

By looking at the upper part of Table \ref{CFPs_dyspro_NewMag}, features similar to those observed in Table \ref{CFPs_cerocene_NewMag} for the cerocene anion are revealed: CASPT2 moderately affects the CFPs while preserving the SOC constant, as expected because the spin–orbit matrix elements evaluated with the CAS wavefunctions remain unchanged. However, the effect is less systematic: if $B_4^0$ and $B_6^0$ are enlarged in absolute values, $B_2^0$ and $B_6^{-5}$ are reduced (in absolute values). If the same simplified reasoning as above is applied, the impact of CASPT2 on the $J$-mixing is expected to be less systematic in this case. Since the ground energy level is composed of the $\ket{J=15/2, M_J=\pm15/2}$ components, it is, in practice, free of $J$-mixing. The first excited Kramers doublet, dominated by the $\ket{J=15/2, M_J=\pm13/2}$ components, is therefore the first to be significantly influenced by this effect. These components mix with the $\ket{J=13/2, M_J=\pm13/2}$ components through the axial CFPs, with the mixing increasing from 3.2\% at the SO-CASSCF level to 4.4\% at the SO-CASPT2 level. For the next Kramers doublet, the $J$-mixing is likewise enhanced. In this case, the $\ket{J=15/2, M_J=\pm11/2}$ components mix with both the $\ket{J=13/2, M_J=\pm11/2}$ and $\ket{J=11/2, M_J=\pm11/2}$ components, with contributions of 1.4\% and 0.4\% at the SO-CASPT2 level, respectively. For the last five Kramers doublets correlating with $\ket{J=15/2}$, CASPT2 reduces the $J$-mixing, in agreement with the expectation above.

\begin{table}[h!]
\caption{Parameter values in cm$^{-1}$ for the $\left[ \text{Dy}(\text{C}_{5}\text{H}_{5})_{2} \right]^{+}$ complex, with the $\left|L,M_L\right>$ or the $\left|J,M_{J}\right>$ basis (full space), based on CAS(9/7) calculations. CFPs with values below 2 cm$^{-1}$ in the \texttt{NewMag} outputs are skipped.\\}
\label{CFPs_dyspro_NewMag}
\centering
\begin{tabular}{|c|cccc|}
\hline
\multicolumn{5}{|c|}{OpenMolcas with associated computational setup (see text)}            \\         
\hline
   & SR-CASSCF & SO-CASSCF & SR-CASPT2 & SO-CASPT2 \\
\hline
$B_2^0$ & 1504.2   & 1504.2   &   1433.9 & 1433.9   \\
$B_4^0$ & 53.2   & 53.2   &   162.1 & 162.1   \\
$B_6^0$ & $-$83.2   & $-$83.2     &  $-$93.1    &   $-$93.1   \\
$B_6^{-5}$ & $-$215.1    & $-$215.1     &  $-$175.5    &     $-$175.5  \\[0.1cm] 
$\zeta{(4f)}$& n/a  & 1974.6    & n/a       &  1974.6  \\[0.1cm] 
MAE & 0.0 & 6.1 & 0.0 & 6.2 \\[0.1cm] 
\hline
\multicolumn{5}{|c|}{ORCA with associated computational setup (see text)}              \\         
\hline
    & SR-CASSCF & SO-CASSCF & SR-NEVPT2 & SO-NEVPT2 \\
\hline
$B_2^0$ & 1573.7   & 1573.7    &   1463.1 &   1463.1  \\
$B_4^0$ & 54.9   & 54.9    &  9.4  &  9.4  \\
$B_6^0$ & $-$83.5   & $-$83.5     &  $-$116.2    & $-$116.2     \\
$B_6^{-5}$ & $-$237.8 & $-$237.8     &  $-$228.9    &   $-$228.9   \\[0.1cm] 
$\zeta{(4f)}$& n/a  & 1931.4    & n/a       & 1931.4   \\[0.1cm] 
MAE & 0.0 & 7.6 & 0.0 & 7.6 \\[0.1cm] 
\hline 
\end{tabular}
\end{table}

It is also interesting to estimate the MAE by limiting our model to given ranks. At the SR-CASPT2 level, the MAE is 249.1 cm$^{-1}$ with only rank-2 operators, 169.6 cm$^{-1}$ with rank-2 + rank-4 operators, 48.6 cm$^{-1}$ with rank 2 + ... + rank-6 operators, 34.1 cm$^{-1}$ with rank-2 + ... + rank-8 operators and eventually 0.0 cm$^{-1}$ with rank-2 + ... + rank-10 operators, to be compared with the \textit{ab initio} spectral width of 1850 cm$^{-1}$ (note that the SOCI spectral witdth is of course much larger because of the SOC when the full $^6$H manifold is considered). Therefore, even if operators up to rank-6 dominate, rank-8 and rank-10 operators are not bland. It is also interesting to see that the first two SR roots are not degenerate anymore starting with the introduction of rank-6 operators. This is due to the occurrence of non-zero $B_k^{-5}$ CFPs (in this coordinate frame, it could have been the $B_k^{5}$ ones if $x$ and $y$ were inverted), which formally breaks the axiality of the system (even if it closely remains axial in practice).

At the SO-CASPT2 level, the MAE is computed based on Equation \ref{re}, meaning that the SOC is also accounted for. Because the SOC is much larger than the CF for lanthanide complexes, here the MAE decreases faster, from 244.4 cm$^{-1}$ with the SOC + rank-2 operators to its minimum value of 6.2 cm$^{-1}$, to be compared with the \textit{ab initio} spectral width of 12506 cm$^{-1}$. If the limit is set to SOC + rank-2 + ... + rank-6 operators, the MAE drops at 14.1 cm$^{-1}$. That is, the MAE becomes more than threefold smaller than with SR-CASPT2 withe the rank-2 + ... + rank-6 operators. Since the CFP values are identical at both levels, this only means that the SOC has a larger impact on the spectrum than the CF, which is obvious here from the comparison of the spectral widths. Because of this, the reconstructed spectrum with the SOC + rank-2 + ... + rank-6 operators is quite a good compromise actually, which qualitatively supports the choices of \texttt{AILFT} (note that the actual values of the CFPs will be discussed in more details below). If the MAE cannot drop to zero at the SOCI level, it is simply because the SOC model with only one parameter (spherical approximation) cannot account for all the fine features of the SOCI matrix. Thus the error is here committed on the SOC model and not on the CF one. A simple extension of the SOC model does not seem to suffice to improve this, therefore, we have retained the one-parameter SOC model at this stage.

Results obtained with ORCA are reported in the bottom part of Table \ref{CFPs_dyspro_NewMag}. Note that the SARC2-DKH-QZVP basis set was used for Dy, and all the other atoms were described with DKH-def2-TZVP basis sets. Overall, the two parts of Table \ref{CFPs_dyspro_NewMag} show a strong resemblance. Naturally, because different basis sets are employed, the SR-CASSCF and SO-CASSCF results already differ to some extent. CASPT2 and NEVPT2 may then introduce further differences. This is particularly apparent for the $B_4^{0}$ term, which is enhanced by CASPT2 but reduced by NEVPT2; the use of different basis sets further impacts the comparison between these results. In the absence of a well-established reference, it is difficult to assess which of the two results is more accurate. The choice is therefore left to the user, who can perform the appropriate methodological tests depending on the system and computational protocol. Overall, these results demonstrate that the implementation is compatible with both OpenMolcas and ORCA, including their respective computational workflows.

Retaining ORCA calculations, it is interesting to compare parameters extracted with \texttt{NewMag} with the ones generated by \texttt{SINGLE\_ANISO} and \texttt{AILFT}.  We start by discussing the \texttt{SINGLE\_ANISO} data (see Table \ref{CFPs_dyspro_comp}, upper part). As for the $f^1$ configuration \cite{NewMag_Sergentu:2026}, the SR-CASSCF \texttt{NewMag} and \texttt{SINGLE\_ANISO} CFPs resemble, here the axial terms are practically identical, we only observe a marginal difference for the $B_6^{-5}$ parameter, and the same is observed at the SR-NEVPT2 level, as expected. Therefore, at the SR level, \texttt{NewMag} reproduces data delivered by \texttt{SINGLE\_ANISO}. At the SOCI level, since \texttt{SINGLE\_ANISO} makes use of the pseudo-$J$ approximation, it does not consider the same number of CFPs as us at the SO-CASSCF and SO-CASPT2 levels (see Tables \ref{ranks} and \ref{ranksANISO}). With \texttt{SINGLE\_ANISO}, the SOC seems to significantly polarize the CFPs, even leading to a sign inversion for $B_4^0$ with the NEVPT2 energies. By construction, \texttt{SINGLE\_ANISO} exactly reproduces the energies of the lowest 8 Kramers doublets that correlate with the $\ket{J=15/2}$ manifold. But for this, phenomenological rank-12 and rank-14 operators need to be introduced, while they are not at the SR level or at the SOCI level with \texttt{NewMag}. To fairly compare the capabilities of \texttt{SINGLE\_ANISO} and  \texttt{NewMag} to reproduce the level energies correlating with $^6H_{15/2}$, we have estimated the MAE obtained on this space, with parameters extracted at the SOCI level, with both codes, by limiting the set of CFPs up to rank-10 only (note that \texttt{NewMag} also includes the SOC). If the \texttt{SINGLE\_ANISO} reconstruction of the model Hamiltonian is limited to rank-10, the MAE committed on the 8 targeted energies is 1.2 cm$^{-1}$ with \texttt{SINGLE\_ANISO}, while it is 9.1 cm$^{-1}$ with \texttt{NewMag}. Thus, it is clear that if \texttt{SINGLE\_ANISO} is better at reproducing those energies, this conclusion is independent from the introduction of the rank-12 and rank-14 operators and it is bound to the targeted space. However, since \texttt{NewMag} targets the full set of $^6H_{15/2}$, $^6H_{13/2}$, \ldots, and $^6H_{5/2}$ manifolds rather than only the $^6H_{15/2}$ manifold, its use may be preferred depending on the research objectives. The two codes are thus complementary, and anyway the users can always choose to focus on the SR parameters even with \texttt{SINGLE\_ANISO}.

\begin{table}[h!]
\caption{Parameter values in cm$^{-1}$ for the $\left[ \text{Dy}(\text{C}_{5}\text{H}_{5})_{2} \right]^{+}$ complex, with the $\left|L,M_L\right>$ or the $\left|J,M_{J}\right>$ basis (full space), based on CAS(9/7) calculations with ORCA. CFPs with values below 2 cm$^{-1}$ are skipped.\\}
\label{CFPs_dyspro_comp}
\centering
\begin{tabular}{|c|cccc|}
\hline
\multicolumn{5}{|c|}{\texttt{SINGLE\_ANISO}}            \\         
\hline
   & SR-CASSCF & SO-CASSCF & SR-NEVPT2 & SO-NEVPT2 \\
\hline
$B_2^0$ & 1573.7   & 1620.6    &   1463.1 &   1500.7  \\
$B_4^0$ & 54.9   & 19.9    &  9.4  &  $-$19.9  \\
$B_6^0$ & $-$83.5   & $-$68.9     &  $-$116.2   &    $-$102.0   \\
$B_6^{-5}$ & $-$241.3 & $-$241.9     &  $-$232.3    &   $-$226.2   \\[0.1cm] 
\hline
\multicolumn{5}{|c|}{\texttt{AILFT}}              \\         
\hline
    & SR-CASSCF & SO-CASSCF & & \\
\hline
$B_2^0$ & 1731.4  & n/a & &  \\
$B_4^0$ & 25.8   & n/a & &  \\
$B_6^0$ & $-$93.8  & n/a & &     \\
$B_6^{-5}$ & $-$220.5  & n/a & & \\[0.1cm] 
$\zeta{(4f)}$& n/a  & 1938.2 & & \\[0.1cm] 
\hline 
\end{tabular}
\end{table}

Results obtained with \texttt{AILFT} are also reported in Table \ref{CFPs_dyspro_comp} (lower part). Since we have only computed 11 SR roots, \texttt{AILFT} was not operative at the NEVPT2 level, therefore such data is not given in Table \ref{CFPs_dyspro_comp}. This is not a mistake but rather a wise choice to ensure that using the exact same SA-CASSCF orbitals with all the three codes that we compare. Qualitatively, \texttt{AILFT} delivers the same picture as \texttt{NewMag}. However, while in the $4f^1$,  $4f^6$, $4f^8$ and $4f^{13}$ configurations the same results were obtained (both the monoelectronic and polyelectronic pictures are identical with only one effective active electron/hole), here we see significant differences in the extracted values. Since the \texttt{NewMag} values allow to map exactly the 2$L$+1 SR energies, while a larger error is done by \texttt{AILFT} due to the choice of parameterization (monoelectronic CF picture, simple model for the bielectronic integrals with as few parameters as possible), the \texttt{NewMag} (or the SR \texttt{SINGLE\_ANISO} values) may in fact be better suited to reproduce (partial) \textit{ab initio} spectra (though \texttt{AILFT} may handle a larger space than the two other codes). Nevertheless, \texttt{AILFT} provides a qualitatively correct description of the CF in a more intuitive form, as diagonalization of the \texttt{AILFT} CF matrix yields one-electron orbital energies. These can provide a useful basis for discussing the CF picture with experimentalists. Thus, \texttt{NewMag} is also complementary with \texttt{AILFT}. Concerning the \texttt{AILFT} SOC constant, which is taken directly from the ORCA output (not reported by \texttt{NewMag}), it is worth noting that the two quantities are defined differently. The \texttt{NewMag} constant is obtained using an ITO procedure based on the same effective Hamiltonian employed to derive the CFPs, whereas \texttt{AILFT} extracts the SOC constant directly from the \textit{ab initio} SOCI matrix. Despite this difference in definition, the two values are relatively close, indicating that the extraction procedure used in \texttt{NewMag} is also consistent with respect to the SOC. 

\noindent \fbox{%
\begin{minipage}{0.46\textwidth}
The current \texttt{NewMag} implementation was showcased with a model dysprosium(III) complex, demonstrating that consistent results can be generated for a polyelectronic case from both OpenMolcas and ORCA calculations. Comparison with the two established codes, \texttt{SINGLE\_ANISO} and \texttt{AILFT}, shows that the present approach is complementary to both and helps bridge the gap between them. It combines some of their respective strengths while enabling additional analyses, particularly of $J$-mixing. It also provides a means of assessing the assumptions underlying these approaches: for example, the relevance of the pseudo-$J$ approximation can be explicitly evaluated for a given system (\textit{cf.} \texttt{SINGLE\_ANISO}), while the magnitude and impact of CFPs of rank 8 and higher can be assessed when applicable (\textit{cf.} \texttt{AILFT}).
\end{minipage}
}

\subsubsection{\label{subsec:Eu-Zn} Application to a real europium(III) complex ($\bm{f^6}$) \protect\\ }

\begin{figure}
    \centering
    \includegraphics[width=1\linewidth]{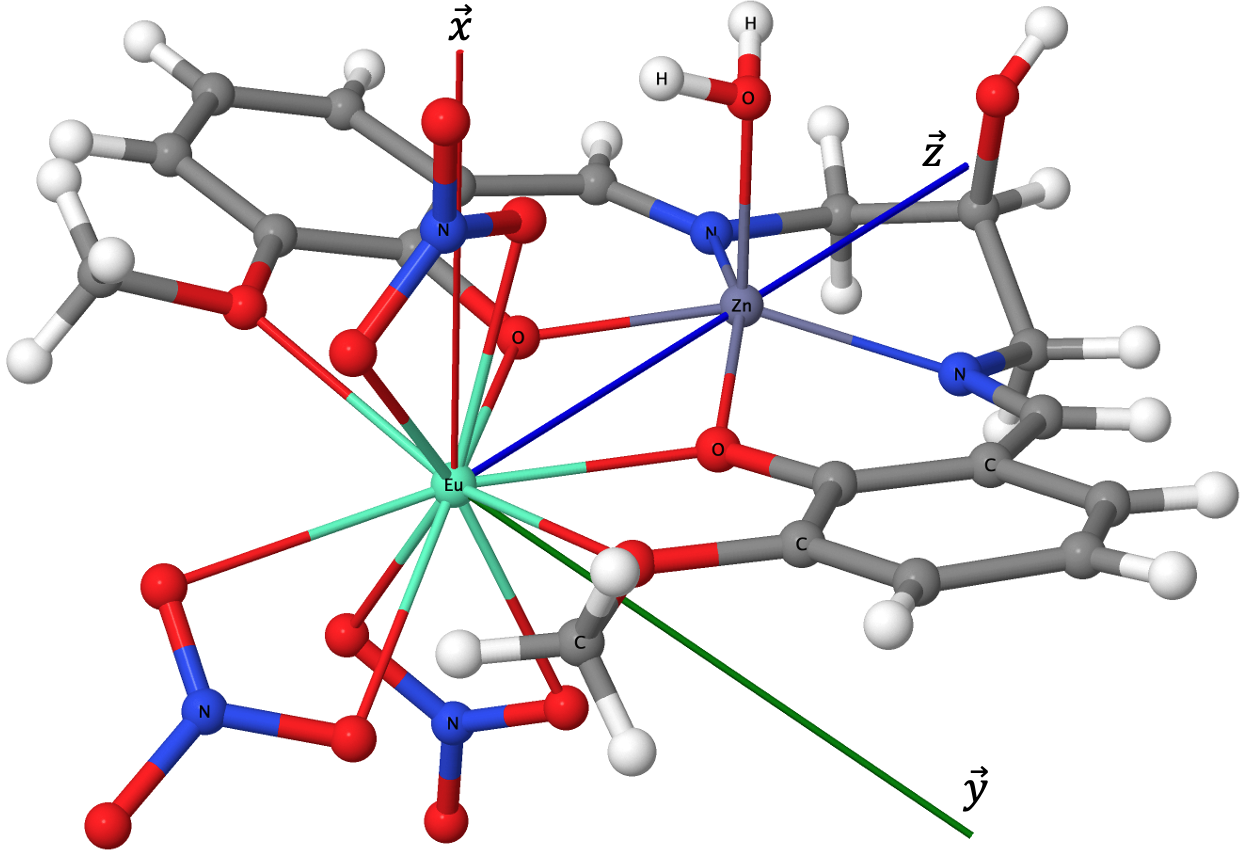}
    \caption{Representation of the $C_{1}$ structure of the considered europium(III) complex \cite{Europium}. The $\vec{z}$ axis is aligned with the Eu--Zn orientation, while the $\vec{x}$ iand $\vec{y}$ axes are arbitrarily chosen. This defines frame (a). Color code: Eu = green, Zn = indigo, O = red, N = blue, C = gray and H = white.}
    \label{euro}
\end{figure}

This section demonstrates that \texttt{NewMag} can extract parameters for realistic, large complexes, handle the extraction of CFPs for the $4f^6$ configuration based on both the SR and SOCI wave functions, and highlight the usefulness of rotationally invariant parameters. For this, we retained a recently reported europium(III) complex (see Figure \ref{euro}), displaying interesting luminescence properties \cite{Europium}, with observed decays to the $^7F_0$, $^7F_1$, $^7F_2$, $^7F_3$ and $^7F_4$ manifolds.

\begin{figure}
    \centering
    \includegraphics[width=1\linewidth]{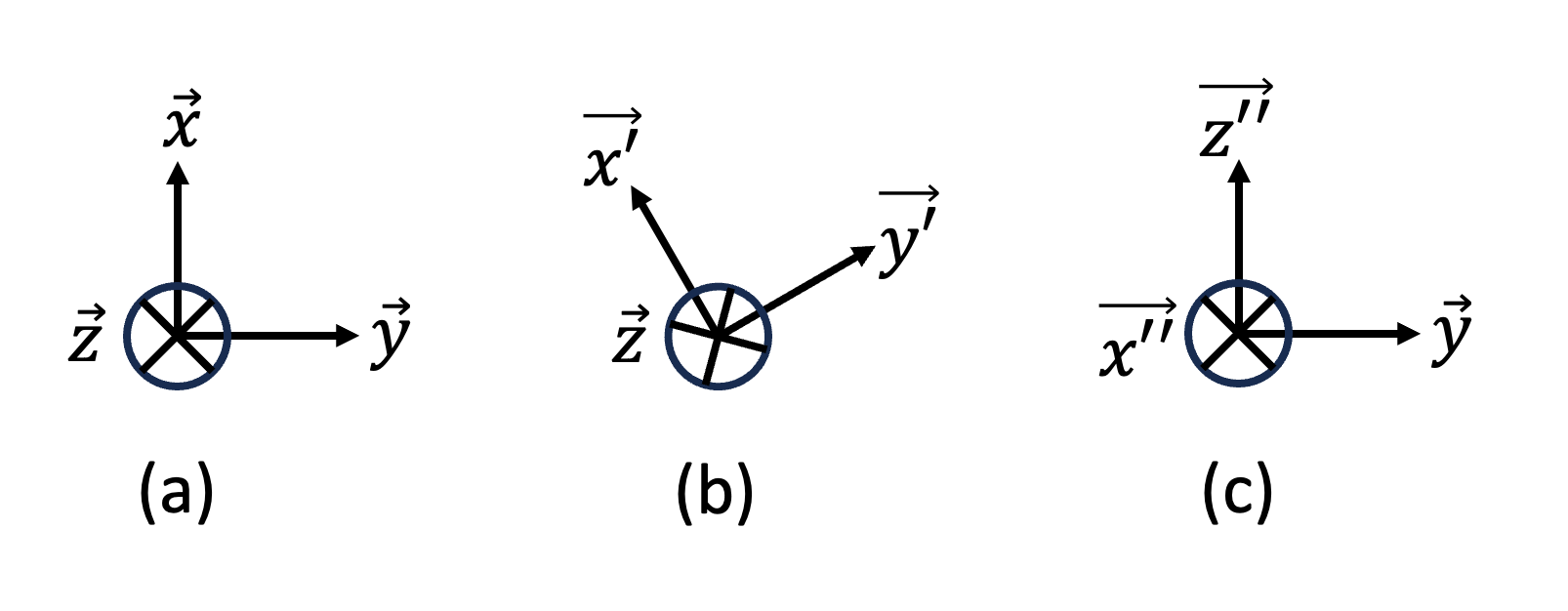}
    \caption{Schematic representation of the three frames used for the considered europium(III) complex: (a) is the initial frame, defined in Figure \ref{euro}, (b) is obtained by applying a rotation around the $\vec{z}$ axis, which leads to the $\vec{x^\prime}$ and $\vec{y^\prime}$ axes, and (c) is obtained from (a) by interverting the $\vec{x}$ and $\vec{z}$ axes, which leads to the $\vec{z^{\prime\prime}}$ and $\vec{x^{\prime\prime}}$ axes.}
    \label{frames}
\end{figure}

Calculations have been performed with ORCA, using the SARC2-DKH-QZVP basis set for Eu, the DKH-def2-TZVP basis set for Zn, and the DKH-def2-SV(P) basis sets for the remaining atoms (O, N, C and H). The 63-atom system belongs to the $C_1$ symmetry point group, meaning that 7 distinct \textit{ab initio} energies are expected at the SR level and 49 at the SOCI level. At the SR-CASSCF and SO-CASSCF levels, 3 coordinate frames were considered (see Figure \ref{frames}), to illustrate how the rotationally-invariant parameters behave, and we have also performed SR-NEVPT2 and SO-NEVPT2 calculations in a selected frame. 

The CFPs extracted by \texttt{NewMag} are of excellent quality, leading to \emph{ab initio} energies with MAE = 0.0 cm$^{-1}$ at the SR levels, and MAE = 0.4-0.5 cm$^{-1}$ at the SOCI levels. Because of the $4f^6$ configuration, operators up to rank-6 are allowed (27 CFPs), enabling the opportunity to test rotationally-invariant parameters without any bias (they are defined based on operators up to rank-6 only). The data are given in Table \ref{CFPs_euro_NewMag}. At first, it is clear that the rank-by-rank contributions to the CF ``strength'' $S$ are identical regardless of the coordinate frame, which is expected (it confirms that ``normalization'' of the CFPs is correctly applied by \texttt{NewMag}). Moreover, $S_2$ is larger than both $S_4$ and $S_6$, indicative of a stronger contribution to the CF ``strength''. This is not equivalent to analyzing the reconstructed model spectra by truncating the CF model at given ranks, $S_2$, $S_4$ and $S_6$ being indicative of the magnitude of the CFPs, and the recontruction indicative of their impact on the spectrum (recall, the CFPs are associated to coefficients in the model matrices, see for instance matrices in our previous paper\cite{NewMag_Sergentu:2026}). For instance, at the SO-NEVPT2 level, the MAE evolves as follows: 53.4 cm$^{-1}$ with the SOC + rank-2 operators, 41.4 cm$^{-1}$ with the SOC + rank-2 + rank 4 operators, and 0.5 cm$^{-1}$ with the SOC + rank-2 + rank 4 + rank-6 operators. Thus the rank-6 operators are more impactful than the rank-4 ones, despite comparable magnitudes ($S_6$ is in fact even smaller than $S_4$). 

\begin{table}[h!]
\caption{Parameter values in cm$^{-1}$ for the considered europium(III) complex, with the $\left|L,M_L\right>$ or the $\left|J,M_{J}\right>$ basis (full space), based on CAS(6/7) calculations with ORCA.\\}
\label{CFPs_euro_NewMag}
\centering
\begin{tabular}{|c|ccc|c|}  
\hline
\multirow{2}{*}{} & \multicolumn{3}{c|}{CASSCF}    & NEVPT2        \\      
\cline{2-5}
   & frame (a) & frame (b) & frame (c) & frame (a)\\
\hline
$S_2$ & 93.5 & 93.5 & 93.4 & 111.6 \\
$S_4$ & 24.0 & 24.0 & 24.0 & 27.8 \\
$S_6$ & 22.9 & 22.9 & 22.9 & 29.7 \\
\hline
$S^0$ & 88.0 & 87.9 & 44.4 & 104.1 \\
$S^1$ & 27.0 & 27.0 & 34.5 & 34.7 \\
$S^2$ & 34.3 & 34.4 & 77.2 & 41.8 \\
$S^3$ & 8.4 & 8.4 & 8.0 & 9.8\\
$S^4$ & 9.8 & 9.8 & 18.1 & 13.4 \\
$S^5$ & 4.0 & 4.0 & 4.6 & 6.0 \\
$S^6$ & 3.4 & 3.4 & 17.1 & 4.3 \\
\hline
$S$ & 57.3 & 57.3 & 57.2 & 68.6 \\
\hline 
\end{tabular}
\end{table}

Finally, it is worth analyzing order-by-order contributions. Switching from frame (a) to frame (b), the $z$ Cartesian axis is left untouched, meaning that all the axial parameters are also identical. For the non-axial parameters, even if individual parameters are changed (for instance, a transfer between the $B_2^{-1}$ and $B_2^{1}$ parameters occurs), $S_1$ is left invariant, by construction, and so on for the higher orders, as stated by Alessandri \textit{et al.} \cite{Riccardo:2018a}. Note that this is in fact another indication that the \texttt{NewMag} implementation is correct. Switching from frame (a) to frame (c), the $z$ Cartesian axis is changed. Consequently, all the individual order-by-order contributions are changed, starting from the axial $S^0$ parameter, but in a way that maintains $S$ if computed with Equation \ref{o}, of course.

Finally, the CASSCF and NEVPT2 CFPs obtained with frame (a) were compared. Apart from $B_4^{-3}$ which is templated by NEVPT2, all the other parameters are enlarged in absolute values. The data in Table \ref{CFPs_euro_NewMag} show that the CF ``strength'' is enhanced by NEVPT2. This can be readily identified from the $S$ parameter computed by \texttt{NewMag}, without having to inspect the 27 individual CFPs. A larger degree of $J$-mixing can therefore be expected at the NEVPT2 level, as illustrated, for example, by the ground energy level, which exhibits 97.8\% $\ket{J=0}$ character at the SO-CASSCF level and 96.9\% at the SO-NEVPT2 level. \\

\noindent \fbox{%
\begin{minipage}{0.46\textwidth}
Following Rudowicz normalisation\cite{Rudowicz:1985} and the work of Alessandri \textit{et al.}\cite{Riccardo:2018a}, we have succesfully implemented rationally-invariant CFPs in \texttt{NewMag}. These indicators may be of interest to quickly highlight features of the CF in given systems. Morever, we are capable of modelling the full space of the $^7F$ manifold of the $4f^6$ configuration, which may be of interest to understand the luminescence properties of compelling europium(III) complexes.
\end{minipage}
}

\subsection{\label{subsec:dn_case} Bonus: Application to d-element systems \protect\\ }

Although the $5d^1$ configuration has already been discussed, this does not imply that \texttt{NewMag} can currently handle all $d^n$ configurations equally successfully. Two prototypical cases are therefore selected to illustrate potential limitations of the present implementation. These examples also help clarify why the approach performs particularly well for lanthanide complexes, while limitations may arise for transition-metal complexes. Actinide complexes are not considered here, as they present additional challenges that would warrant a dedicated study\cite{preprint}.

\subsubsection{\label{subsec:Mn} Case 1: A model manganese(III) complex  ($\bm{d^4}$) \protect\\ }

Transition metal complexes are usually distinct from lanthanide complexes in the sense that the CF is much stronger than the SOC, and that the CF picture may be challenging. With a strong enough CF, spin transitions may even occur, meaning that the ground state may not follow Hund's rule. Without tackling those extreme cases, we may still define situations to push \texttt{NewMag} to its limits. 

A first application concerns octahedral manganese(III) complexes. It is known that such complexes are not stable, since the Jahn-Teller effect should trigger an axial elongation (or even a compression \cite{Fackler:1974}). However, from a theoretical viewpoint, the octahedral situation is quite interesting. The CF splits in this case the $^5D$ free ion term into two SR states, $^5E_g$ and $^5T_{2g}$. Despite the orbital degeneracy of the ground SR state, the components of $^5E_g$ cannot be coupled by the SOC, but their coupling with the $^5T_{2g}$ ones leads to 10 first energy levels, clearly correlating with $^5E_g$, with degeneracies of 1, 3, 2, 3 and 1 in ascending energetic order \cite{Abragam:1970a}. If one further adds the lowest SR triplet spin components, that is the components of $^3T_{1g}$ (which correlates with $^3H$ of the reference free ion), this 1, 3, 2, 3 and 1 pattern is maintained but the energy spacings are enhanced \cite{Maurice:Mn}. Since SOCs with other spin state components cannot be handled by \texttt{NewMag}, this must affect the quality of the model spectrum, and this is exactly what we aimed at assessing.

\begin{figure}
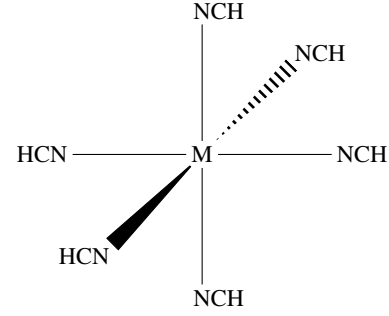

\chemfig{[0,2]HCN-M(<:[1]NCH)(-[2]NCH)(<[5]HCN)(-[6]NCH)-NCH}
\caption{Schematic representation of the $O_h$ structure of [M(NCH)$_6$]$^{3+}$, where M is either Mn or Ni.} 
\label{oh}
\end{figure}

We have considered a model [Mn(NCH)$_6$]$^{3+}$ complex of octahedral symmetry (see Figure \ref{oh}), using the same geometry reported by the previous publication\cite{Maurice:Mn}. Calculations were performed with OpenMolcas, which is particularly suitable for this demonstration because the SA-CASSCF orbitals are constructed separately for each spin block. Consequently, enlarging the SOCI space to include additional spin-state components is not expected to affect the extracted CFPs, in contrast to what would be observed with ORCA. This provides an opportunity to examine the behavior of \texttt{NewMag} when ``second-order'' SOC effects are involved. The order of a perturbative effect is intrinsically linked to the choice of model (or reference) space. In \texttt{NewMag}, the full $^5D$ manifold is included in the model space, so the couplings between the $^5E_g$ and $^5T_{2g}$ components arise at first order. In contrast, in the previously mentioned publication by one of the authors \cite{Maurice:Mn}, the model space comprised only the $^5E_g$ components, and the same $^5E_g$--$^5T_{2g}$ couplings were therefore described as ``second-order'' couplings. The underlying physical couplings are nevertheless exactly the same.

Three sets of calculations were performed, with the SA-CASSCF and SO-CASSCF methods, with the full quintet manifold (the target of \texttt{NewMag}), this manifold plus $^3T_{1g}$, and the full quintet manifold plus 11 triplets (correlating with $^3H$ of the reference free ion). Note that with the current implementation, it is crucial to include the high-spin states first in the input (for both OpenMolcas and ORCA), since \texttt{NewMag} will assume that the targeted 2$L$+1 states appear first in the output. The ANO-RCC-VTZP basis set was used for Mn and the ANO-RCC-VDZP ones for the remaining atoms (N, C and H). Results are presented in Table \ref{CFPs_mn_NewMag}. 

\begin{table}[h!]
\caption{Parameter values in cm$^{-1}$ for the [Mn(NCH)$_6$]$^{3+}$ complex, with the $\left|L,M_L\right>$ or the $\left|J,M_{J}\right>$ basis (full space), based on CAS(4/5) calculations with OpenMolcas. CFPs with values below 1 cm$^{-1}$ in the \texttt{NewMag} outputs are skipped.\\}
\label{CFPs_mn_NewMag}
\centering
\begin{tabular}{|c|cc|}
\hline
\multicolumn{3}{|c|}{With the full quintet manifold ($^5D$)}            \\         
\hline
   & SR-CASSCF & SO-CASSCF  \\
\hline
$B_4^0$ & 3710.3   & 3710.3     \\
$B_4^4$ & 18551.4    & 18551.4        \\[0.1cm] 
$\zeta{(3d)}$& n/a  & 355.6           \\[0.1cm] 
MAE & 0.0 & 1.3 \\[0.1cm] 
\hline
\multicolumn{3}{|c|}{With the full quintet manifold ($^5D$) + 3 triplets}              \\         
\hline
    & SR-CASSCF & SO-CASSCF  \\
\hline
$B_4^0$ &  3710.3  &    3712.8  \\
$B_4^4$ &    18551.4 &    18564.2     \\[0.1cm] 
$\zeta{(3d)}$ & n/a  &      355.7    \\[0.1cm] 
MAE & 0.0 &  10.1  \\[0.1cm]  
\hline 
\multicolumn{3}{|c|}{With the full quintet manifold ($^5D$) + 11 triplets ($^3H$)}              \\         
\hline
    & SR-CASSCF & SO-CASSCF  \\
\hline
$B_4^0$ &  3710.3 &   3710.1   \\
$B_4^4$ &  18551.4   &    18550.7     \\[0.1cm] 
$\zeta{(3d)}$ & n/a  &  356.2       \\[0.1cm] 
MAE & 0.0 & 11.2  \\[0.1cm]  
\hline 
\end{tabular}
\end{table}

We start by analyzing the data obtained with the complete quintet manifold. In the $O_h$ symmetry point group, for the $3d^4$ configuration, only two CF parameters are at play, $B_4^0$ and $B_4^4$. As already known \cite{Abragam:1970a}, in this case, $B_4^4$ = 5 $B_4^0$. This is exactly what is obtained with \texttt{NewMag}. At the SR-CASSCF level, a perfect reproduction of the \textit{ab initio} spectrum is observed (MAE = 0.0 cm$^{-1}$). At the SOCI level, the model is also quite accurate, with an MAE of 1.3 cm$^{-1}$. As expected, the present CF model correctly reproduce the previously mentioned 1, 3, 2, 3, 1 degeneracy pattern for the lowest 10 energy levels (this is necessarily the case, as it follows directly from symmetry).

Since the SA-CASSCF orbitals are maintained for the quintets in the other two sets of calculations, the SR-CASSCF remains identical. At the SOCI level, however, significant differences emerge. The present model does not account for the second-order SOCs and the error committed is enlarged by a factor of magnitude. Most of this discrepancy originates from the $^3T_{1g}$ components, which explains the similar MAEs obtained for the two corresponding calculations. Because the model defined by Equation \ref{re} is no longer fully adequate for describing the \textit{ab initio} effective Hamiltonian in this case, the extracted CFPs become biased. In particular, the inclusion of SOC appears to induce a polarization of the CFPs, which is undesirable at the SOCI level considered here.\\

Of course, one may wonder if such a situation is susceptible to occur in lanthanide complexes. If SOCs with lower spin-state components occur, it should result in ``$S$-mixing''. A recent study by Zhang and Yang \cite{spinmix} attempted to tackle this issue under strict axial symmetry. In their study, the authors showed that the trivalent Ln(COT)$^+$ (Ln = Ce, Pr and Nd) complexes ($f^n$ electronic configurations, with $n$ = 1--3) displayed no to little $S$-mixing, meaning that with a similar computational setup (\textit{i.e.} with the same configuration interaction space) \texttt{NewMag} should work quite well at the SOCI level. However, in the divalent Ln(COT) complexes ($f^n6s^1$ electronic configurations, with $n$ = 1--3), important $S$-mixings may be observed. Since \texttt{NewMag} has not been built for this, the values of the CFPs that would in this case be extracted at the SOCI level would have to be taken with care. Anyway, the extracted parameters at the SR level, both in the divalent and trivalent cases, should be correct, exactly in the same vein as the octahedral Mn(III) case.

\noindent \fbox{%
\begin{minipage}{0.46\textwidth}
When the SOCI calculation only accounts for SOCs within the model space employed by \texttt{NewMag} (first-order SOCs), the code is well suited to extract CFPs at the SOCI level, as demonstrated for the $5d^1$ configuration in Section \ref{subsec:fluo_Ce}. Otherwise, the extracted values may be biased, compromising the quality of the reconstructed model spectrum. In general, we expect no issue with trivalent lanthanide complexes, while the case of divalent lanthanide complexes may in fact be more challenging.
\end{minipage}
}

\subsubsection{\label{subsec:Ni} Case 2: A model nickel(II) complex  ($\bm{d^8}$) \protect\\ }

High-spin nickel(II) complexes are typically discussed in terms of zero-field splitting (ZFS), which reflects spin anisotropy when the orbital momentum is quenched. Here, however, the discussion is restricted to the isotropic, octahedral case (see Figure \ref{oh}) and therefore do not consider ZFS. A computational setup similar to that used for the manganese(III) complex was employed here, and likewise started from a previously reported structure of [Ni(NCH)$_6$]$^{2+}$ \cite{Chap_O}. It is important to recall that the objective here is to deliberately challenge \texttt{NewMag} already at the SR level. 

\begin{figure}
    \centering
    \includegraphics[width=0.95\linewidth]{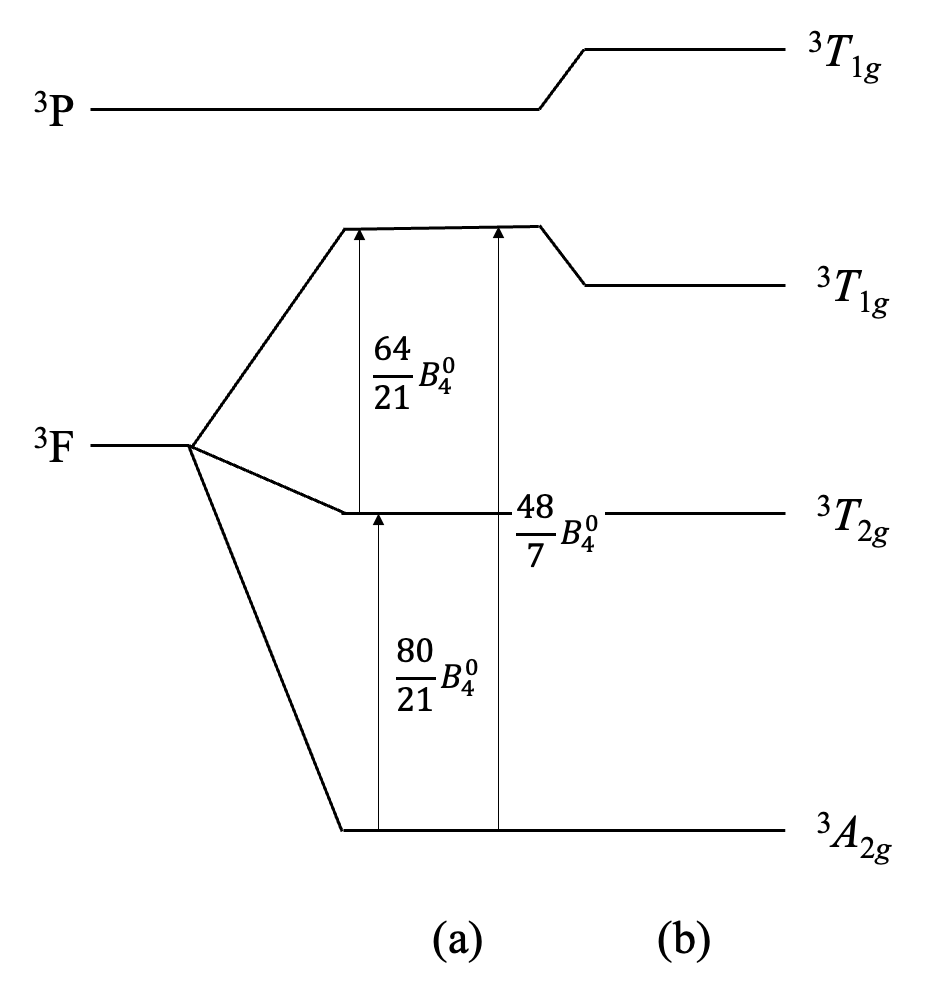}
    \caption{Scheme of the high-spin SR states of $d^8$ configuration. The CF Hamiltonian perfectly described the energies of the 2$L$+1 states of the model space in the true CF picture (a), for which no coupling between the two $^3T_{1g}$ states is possible, unlike in the true LF picture (b), for which this coupling occurs and alters the spectrum.}
    \label{NiOh}
\end{figure}

Within CF theory, the reference free ion $^3F$ term splits into 3 SR states in the $O_h$ symmetry point group with well defined energy spacings (see Figure \ref{NiOh}) \cite{Abragam:1970a, Chap_anal}. To compare with previous references, recall that $B_4$ = $B_4^0$ and that the $\left< L|| \beta ||L\right>$ prefactor, 2/315, was not applied there\cite{Abragam:1970a, Chap_anal} while it has been applied here for consistency with the \texttt{NewMag} implementation. If the energy of $^3A_{2g}$ is set at zero, the following relation is verified:

\begin{equation}
\frac{E(^3T_{1g})}{E(^3T_{2g})}=\frac{9}{5}=1.8
\end{equation}

\noindent If a point-charge model is built (not shown), this ratio is perfectly respected by the SA-CASSCF method, as expected. However, this is not the case for the [Ni(NCH)$_6$]$^{2+}$ model complex, since $E(^3T_{1g})$ = 15395.9 cm$^{-1}$ and $E(^3T_{2g})$ = 9018.1 cm$^{-1}$, leading to a ratio of 1.7 instead of 1.8. Clearly, the ratio is not reproduced by the present CF model. This deviation emerges in a ligand-field picture, where the electron-electron term enables the coupling of the first $^3T_{1g}$ with the other  $3T_{1g}$ of the $d^8$ configuration [the one which correlates with $^3P$ of the reference free ion (see Figure \ref{NiOh})]. Since the $^3A_{2g}$ and $^3T_{2g}$ SR roots are not affected by this coupling, we may use $E(^3T_{2g})$ = 9018.1 cm$^{-1}$ to correctly extract $B_4^0$ and $B_4^4$. With the expression for the energy difference between those two states reported in Figure \ref{NiOh}, the following CFPs should be obtained: $B_4^0$ = 2367.3 cm$^{-1}$ and $B_4^4$ = 5 $B_4^0$ = 11836.3 cm$^{-1}$. Clearly, the values obtained with \texttt{NewMag} differ significantly, and the quality of the model spectrum is also quite poor (see Table \ref{CFPs_ni_NewMag}).

\begin{table}[h!]
\caption{Parameter values in cm$^{-1}$ for the [Ni(NCH)$_6$]$^{2+}$ complex, with the $\left|L,M_L\right>$ or the $\left|J,M_{J}\right>$ basis (full space), based on CAS(8/5) calculations with OpenMolcas. CFPs with values below 0.05 cm$^{-1}$ in the \texttt{NewMag} outputs are skipped.\\}
\label{CFPs_ni_NewMag}
\centering
\begin{tabular}{|c|cc|}  
\hline
   & SR-CASSCF & SO-CASSCF  \\
\hline
$B_4^0$ & 2217.5   & 2217.5     \\
$B_4^4$ & 11087.5   & 11087.5        \\[0.1cm] 
$\zeta{(3d)}$& n/a  & 637.7          \\[0.1cm] 
MAE & 326.0 & 327.0 \\[0.1cm] 
\hline
\end{tabular}
\end{table}

\noindent \fbox{%
\begin{minipage}{0.46\textwidth}
\texttt{NewMag} excels when the LF does not induce significant mixing between the SR states belonging to its model space and those outside it. When this condition is not fulfilled, the extraction of the CF model based on the restricted model space targeted by \texttt{NewMag} should be interpreted with caution. However, in specific cases where the full high-spin manifold is included in the model space, \textit{e.g.}, for the $d^1$ configuration (see Section \ref{subsec:fluo_Ce}) and, at least at the SR level, for the $d^4$ configuration (see Section \ref{subsec:Mn}), \texttt{NewMag} remains fully appropriate. It is also worth returning to our original target, lanthanide complexes. Owing to the core-like nature of the $4f$ orbitals, these systems are expected to be particularly well suited to a pure CF description, which is consistent with the successful treatment of all the lanthanide cases reported here.
\end{minipage}
}

\section{\label{sec:conclusion}Concluding remarks\protect\\ }

CF theory is well established, and so is the extraction of CFPs from relativistic and multiconfigurational \textit{ab initio} calculations. \texttt{NewMag} was designed as a complementary approach to two pioneering codes in the field, namely \texttt{SINGLE\_ANISO} and \texttt{AILFT}. Whereas these codes employ elegant shortcuts to address efficiently a wide range of complexes of interest to the community, we deliberately adopted a more elaborate workflow with two main advantages: (i) to retain explicit information on the nature of the many-electron states throughout the procedure, and (ii) to provides a means of assessing the validity of some of the assumptions underlying these approaches, which is not possible within their respective workflows by design. Thus, beyond its appeal from a quantum-chemical perspective, \texttt{NewMag} provides a useful complementary tool to these established approaches.

Presently, \texttt{NewMag} can handle OpenMolcas and ORCA outputs, processing SR-CASSCF, SO-CASSCF, SR-CASPT2, SO-CASPT2, SR-NEVPT2 and/or SO-NEVPT2 calculations with and beyond minimal active spaces, and performs very well for any trivalent lanthanide complex of the $f^1$--$f^6$ and $f^8$--$f^{13}$ configurations. 

Though technically nothing prohibits to tackle transition metal complexes with \texttt{NewMag}, it should be understood that it has intrinsic limitations, meaning that it may not be generally consistent and accurate in these cases, simply because it neglects couplings between the SR states of the model space with external states, and it also neglects second-order SOCs for instance with states of different spin multiplicities. Note that these issues are expected to be far less important in trivalent lanthanide complexes, which explains why \texttt{NewMag} is generally consistent and accurate there. 

Likewise, actinide complexes may pose additional difficulties to \texttt{NewMag}, by combining features of both lanthanide and transition metal complexes. Therefore, we have not reported any case in this article, leaving it as a perspective of this work.

Further conclusions were given at the end of each case study and are not repeated here. The interested reader may thus also consult those conclusions as a complement.

\section*{Conflict of Interest}
\noindent The authors have no conflicts to disclose.

\begin{acknowledgments}
The authors thank Rémi Marchal for the administration of the HPC cluster of ISCR as well as for precious technical assistance. D.-C.S. acknowledges mobility funding provided through the Romania–France bilateral research program, supported by the Romanian National Authority for Scientific Research and Innovation (UEFISCDI), Project No. PN-IV-P8-8.3-PM-RO-FR-2024-00-22. G.D.-R., B.L.G., and R.M. acknowledge the “PHC Brancusi” program (Project No. 51686YA), funded by the French Ministry for Europe and Foreign Affairs, the French Ministry for Higher Education and Research, and the Ministry of Research, Innovation and Digitalization (M.C.I.D.). Additional support by the ANR (Contract No. ANR-23-PETQ-0007) is also acknowledged.
\end{acknowledgments}

\section*{Data Availability Statement}

The data that support the findings of this study are available within the article and at the NewMag code repository, which is publicly available on GitHub at
https://github.com/clausserg/newmag.git. The examples folder in this repository contains all raw output from the \textit{ab initio} calculations reported in this paper as well as the outputs of \texttt{NewMag}. Note that the employed XYZ coordinates are directly available in those outputs. Additional documentation concerning the code and its use is also provided (flowcharts, description of keywords, examples).


\section*{REFERENCES}
\bibliography{references}

\end{document}